\documentclass[prb,aps,twocolumn,showpacs,floats]{revtex4}
\usepackage{graphicx}

\begin{document}

\title{Initial correlations in nonequilibrium Falicov-Kimball model}

\author{Minh-Tien Tran}
\affiliation{Asia Pacific Center for Theoretical Physics, POSTECH, Pohang, Republic of Korea and \\
Institute of Physics and Electronics, Vietnamese Academy of Science and Technology,
10 Dao Tan, Hanoi, Vietnam.
}


\begin{abstract}
The Keldysh boundary problem in a nonequilibrium Falicov-Kimball model in infinite dimensions 
is studied within the truncated and self-consistent perturbation theories, and the dynamical 
mean-field theory. Within the model the system is started in equilibrium, and later 
a uniform electric field is turned on. 
The Kadanoff-Baym-Wagner equations for the nonequilibrium Green functions are derived,
and numerically solved. The contributions of initial correlations are studied by monitoring 
the system evolution. It is found that the initial correlations are essential for establishing 
full electron correlations of the system and independent on the starting time of preparing 
the system in equilibrium.
By examining the contributions of the initial correlations to the electric current
and the double occupation, we find that the contributions are small in relation to
the total value of those physical quantities when the interaction is weak, and
significantly increase when the interaction is strong. 
 The neglect of initial correlations may cause artifacts in 
 the nonequilibrium properties of the system, especially in the strong 
interaction case.
    
\end{abstract}

\maketitle

\section{Introduction}
The theoretical description of the physical properties of nonequilibrium
correlated electron systems is an important problem in condensed matter
physics. In transport processes
often correlated electron systems are driven out of equilibrium
by switching on external fields. The systems can be
also out of equilibrium by suddenly changing their parameters. Nonequilibrium
correlated electron systems, which can be realized in many experiments,  
may have unusual and interesting properties. One such system is the quantum dot
attached to two leads through two tunnel junctions.\cite{Kastner,Kouwenhoven,Simmel}
The conductance of the dot reveals a nonequilibrium Kondo effect. Other examples
are the effects of electron correlations on the nonlinear current-voltage 
characteristics.\cite{Ogasawara,Taguchi} Recently, experiments with ultracold
atomic gases have made it possible to prepare initial state to a rapid change 
of system parameters, and observed remarkable subsequent dynamics as a collapse
and revival of the initial phase.\cite{Greiner,Kinoshita}

The many-body formalism for nonequilibrium systems was developed by
many people including Kubo,\cite{Kubo} Schwinger,\cite{Schwinger} Kadanoff and
Baym,\cite{Kadanoff} Keldysh\cite{Keldysh} 
(see also Ref.~\onlinecite{Danielewicz,Rammer} for references). In particular,
Kadanoff and Baym constructed a system of equations for the nonequilibrium
Green functions.\cite{Kadanoff} Parallel to this development, Keldysh also derived
a perturbation theory for the nonequilibrium Green functions.\cite{Keldysh} 
Like the Feynman perturbation theory for equilibrium systems,\cite{Mahan}
the Keldysh nonequilibrium perturbation theory is based on the assumption
of an adiabatic switching on of the many-body interactions.  The assumption
is necessary for the application of the Wick theorem, which requires a quadratic
form of the system Hamiltonian at the initial preparation of the system. 
While the assumption is exactly proved in scattering theory,\cite{Mahan} 
its application
to nonequilibrium many-body systems imposes restrictions.\cite{Fujita,Hall,Wagner}
It turns out that the assumption corresponds to the neglect of the so-called initial
correlations.\cite{Fujita,Hall,Wagner} 
Despite the neglect of the initial correlations the Keldysh theory is
widely used in the study of nonequilibrium systems.
Wagner unified the Feynman, Matsubara and Keldysh perturbation
theories into a single many-body formalism in which neither
a special form of the Hamiltonian at the initial time nor subsequent time
development of the system are restricted.\cite{Wagner} He introduced a matrix
representation for the contour-ordered Green function and derived the
Kadanoff-Baym equations for the nonequilibrium Green functions. In the
Kadanoff-Baym-Wagner formalism the initial correlations are fully taken into account.   
While the nonequilibrium formalism is well established, the role of the initial
correlations is less attended, especially for nonequilibrium strongly correlated
electron systems. In particular, it would be desirable to test 
whether the initial correlations which
are neglected in the Keldysh formalism are negligible or not. The difficulties arising  
in the study of the initial correlations are mostly due to the lack of
getting the exact solutions of nonequilibrium correlated electron systems.

In the last decade, the dynamical mean-field theory (DMFT) was 
developed.\cite{Metzner,GKKR} In the equilibrium case the theory is widely
and successfully applied to study strongly correlated electron systems.
The DMFT gives the exact solutions in infinite dimensions. Recently,
a version of the DMFT for nonequilibrium systems was developed.\cite{Freericks}
The nonequilibrium dynamical mean-field theory (NEDMFT) is formally 
formulated on the same basis as of the equilibrium DMFT. Like the 
equilibrium case, in infinite
dimensions the self energy of nonequilibrium systems becomes a local function
in space. As a consequence, it can be self-consistently determined by mapping
the lattice problem onto an effective problem of a single site embedded in 
a self-consistent effective medium. 
When the self-consistent equations are solved, the nonequilibrium Green functions
are obtained and various physical quantities can be calculated.

The aim of the present paper is twofold. First, we study the contributions
of the initial correlations in a nonequilibrium correlated electron system. 
The initial correlations are studied within  truncated
and self-consistent perturbation theories as well as within the NEDMFT. The Keldysh
perturbation theory usually argues for the neglect of the initial correlations.
However, in the present paper the results obtained within the truncated and self-consistent perturbation
theories show that the initial correlations are always finite even when the initial time
is in the remote past limit. In the infinite dimension limit the initial correlations 
can also be obtained exactly since in this limit the NEDMFT gives the exact solutions. 
In the such way one can find under what circumstance the Keldysh formalism is safely applied to
nonequilibrium correlated electron systems. The second aim of the present paper is to derive 
the Kadanoff-Baym-Wagner equations for the nonequilibrium Green functions within the
NEDMFT. These equations are an alternative to the original NEDMFT equations for the
contour-ordered Green function.\cite{Freericks} Within the  Kadanoff-Baym-Wagner
formalism the nonequilibrium Green functions clearly satisfy their 
boundary conditions. The Kadanoff-Baym-Wagner formalism already includes the Keldysh
formalism as its part, and it is suitable to study the initial correlations.
In this paper we will examine the initial correlations
within a Keldysh boundary problem. The problem works in a
system which is first started in equilibrium and then is driven 
out of equilibrium by turning on of an external field.
The model which we adopt to describe the system is a nonequilibrium Falicov-Kimball
model. The Falicov-Kimball model (FKM) was first introduced for modelling a
metal-insulator transition in equilibrium.\cite{Falicov} The model is one of
the simplest models for strongly correlated electron systems. The
FKM describes conduction electrons interacting via a
repulsive contact potential with localized electrons. It can be
viewed as a simplified Hubbard model where electrons with down spin are
frozen and do not hop. Much progress has been made on solving this model 
in both exact and approximation ways, where all properties of the 
conduction electrons in equilibrium are well known.\cite{Kennedy,Lemanski,FreericksZlatic}
In equilibrium the FKM describes a metal-insulator transition for the
homogeneous phase.\cite{Kennedy,Lemanski,FreericksZlatic} The
Coulomb interaction is divided into two ranges: the weak interaction range, when the interaction strength
is smaller than the half bare bandwidth, and strong interaction one otherwise.
For weak interactions the system is metallic, and for strong ones the system is insulator.
The system is driven out of equilibrium by
a constant electric field. The electric field is switched on at a some time after
the initial preparation of the system in equilibrium. This nonequilibrium FKM was introduced
by Freericks {\em et al} in the study of the Bloch oscillations in the electric current
within the NEDMFT.\cite{Freericks} In the present paper we derive the 
Kadanoff-Baym-Wagner equations of the NEDMFT for the nonequilibrium FKM, and calculate
the contributions of the initial correlations to the electric current and the double
occupation. It is found that the contributions of the initial correlations to 
the electric current and the double
occupation
are small in relation to the full value of those physical quantities 
in the weak interaction case, and significantly increases 
in the strong interaction case.
However, without the initial correlations
the system cannot restore full electron correlations even before the turning on of
the electric field. The neglect of the initial correlations may cause artifacts in 
the nonequilibrium properties of the system.

The paper is organized as follows. In Sec.~II we present the Kadanoff-Baym-Wagner 
nonequilibrium formalism, and describe the nonequilibrium FKM. In the next three
sections we present the studies of the model within the truncated and self-consistent 
perturbation theories, and the NEDMFT. The last section is the conclusion. 

\section{Formalism and model}
\begin{figure}[b]
\includegraphics[width=0.42\textwidth]{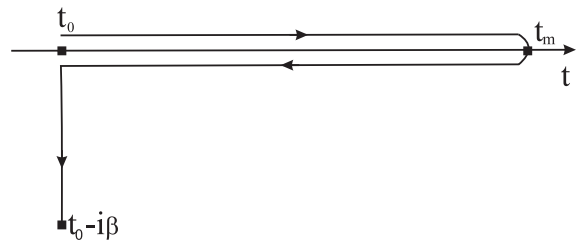}
\caption{Kadanoff-Baym contour for the two-time Green functions in nonequilibrium.}
\label{fig1}
\end{figure}
We consider the Keldysh boundary problem in a nonequilibrium system which is 
first prepared in equilibrium, and then is driven out of equilibrium by switching on of an external field or by a sudden change of its parameters. 
Specifically, at an initial time
$t_0$ the system is prepared in equilibrium which is defined by the 
equilibrium Hamiltonian $H_{\text{eq}}$ and temperature $1/\beta$, and
at time $t=0$ ($t_0 < 0$) an external field is switched on or its parameters are 
suddenly changed. Usually,
in nonequilibrium systems the time translational invariance is not valid, 
and the Green functions, which are employed
for studying the physical properties of the systems,
depend on the two time variables. The nonequilibrium 
formalism works with the so-called contour-ordered Green function which is defined 
for the time variables on the Kadanoff-Baym contour.\cite{Kadanoff,Keldysh,Rammer,Danielewicz}
The Kadanoff-Baym contour is shown on Fig.~\ref{fig1}. The contour starts in the initial
time $t_0$, runs out to maximal time $t_m$, then returns to the initial time, and finally
moves parallel to the negative imaginary axis a distance $\beta$. At the initial time
$t_0$ the system is always in equilibrium.
The Kadanoff-Baym
contour is suitable for deriving the Dyson equation for the contour-ordered 
Green function. Keldysh also introduced a similar contour for the contour-ordered
Green function.\cite{Keldysh} The Keldysh contour  is basically the same as the 
Kadanoff-Baym contour, but it neglects the last contour branch 
parallel to the imaginary
axis and limits $t_0$ to minus infinity. The neglect of the last branch of the contour
corresponds to the neglect of initial correlations.   
The contour-ordered Green function is defined by 
\begin{eqnarray}
G^{c}(i,j | \bar{t},\bar{t}') &=& 
-i \big\langle \mathcal{T}_{c} c_{i}(\bar{t}) c^{\dagger}_{j}(\bar{t}')
\big\rangle  \nonumber \\
&=& - i \theta^{c}(\bar{t},\bar{t}')
\big\langle c_{i}(\bar{t}) c^{\dagger}_{j}(\bar{t}')
\big\rangle  \nonumber \\
&& + i \theta^{c}(\bar{t}',\bar{t})
\big\langle c^{\dagger}_{j}(\bar{t}') c_{i}(\bar{t}) \big\rangle , \label{gc2}  
\end{eqnarray}
where $c^{\dagger}_i$ ($c_i$) are the creation (annihilation) operators for electrons
at site $i$. The time evolution of the operators on the Kadanoff-Baym contour is 
defined in the Heisenberg picture.
$\mathcal{T}_c$ is the time ordering 
on the Kadanoff-Baym contour and it is defined via the contour step function $\theta^c(\bar{t},\bar{t}')$. $\theta^c(\bar{t},\bar{t}')$ equals to $1$ if $\bar{t}$ lies after $\bar{t}'$ on the contour, and it equals to $0$ otherwise. The averages in Eq.~(\ref{gc2}) are the statistical
average over the equilibrium Hamiltonian $H_{\text{eq}}$ at temperature $1/\beta$.
The Kadanoff-Baym contour consists of three time branches: the first branch
is chronological, the second one is antichronological, and the last one is parallel
to the imaginary axis.
Thus, we can represent 
the contour-ordered Green function by a $3\times 3$ matrix 
$G^{\alpha\gamma}(\bar{t},\bar{t}')$, where $\bar{t}$ locates on the $\alpha$-th branch,
and $\bar{t}'$ locates on the $\gamma$-th branch. In the such way, the contour-ordered 
Green function has nine component Green functions, however they are not independent.
Wagner reduced the matrix representation of the contour-ordered Green function 
to a matrix form of six  component Green functions, and five of them are independent.\cite{Wagner} 
The Wagner matrix representation for the contour-ordered Green function can be written as 
follows\cite{Wagner}
\begin{equation}
\widehat{G} =
\left(
\begin{array}{ccc}
G^{R} & G^{K} & \sqrt{2} G^{\rceil} \\
0 & G^{A} & 0 \\
0 & \sqrt{2} G^{\lceil} & G^{M} 
\end{array}
\right) , \label{wagn}
\end{equation}
where 
\begin{eqnarray*}
G^{R}(t,t') &=& G^{11}(t,t')-G^{12}(t,t') \\
&=& -i \theta(t-t') \big\langle \{c_i(t),c^{\dagger}_j(t')\}\big\rangle , \\
G^{A}(t,t') &=& G^{12}(t,t')-G^{22}(t,t') \\
&=& i \theta(t'-t) \big\langle \{c_i(t),c^{\dagger}_j(t')\}\big\rangle , \\
G^{K}(t,t') &=& G^{12}(t,t')+G^{21}(t,t') \\
&=& -i \big\langle [c_i(t),c^{\dagger}_j(t')]\big\rangle , \\
G^{M}(\tau,\tau') &=& G^{33}(t_0-i\tau,t_0-i\tau') \\
&=& -i \big\langle 
\mathcal{T}_{\tau} c_i(t_0-i\tau) c^{\dagger}_j(t_0-i\tau') \big\rangle , \\
G^{\rceil}(t,\tau') &=& G^{13}(t,t_0-i\tau') \\
&=& i \big\langle c^{\dagger}_j(t_0-i\tau') c_i(t)  \big\rangle , \\
G^{\lceil}(\tau,t') &=& G^{31}(t_0-i\tau,t') \\
&=& - i \big\langle c_i(t_0-i\tau) c^{\dagger}_j(t') \big\rangle ,
\end{eqnarray*}
where $t$, $t'$ are real times, and $0\leq \tau, \tau' \leq \beta$.
In the above equations we have used the commutator symbol 
$[A,B ]=AB-BA$, and the anticommutator symbol $\{A,B \}=AB+BA$. 
$G^{R(A)}$ are the retarded (advanced) Green function, and 
$G^{K}$ is the Keldysh 
Green function. These Green functions are defined totally on the real time axis. 
$G^{M}$ is the Matsubara Green function and is defined 
on the imaginary time branch of the Kadanoff-Baym contour. Note that  the above definition of
the Matsubara Green function differs from the standard one by 
factor $i$.\cite{Mahan}
The Green functions
$G^{\rceil}$ and $G^{\lceil}$ have one time variable on the real time axis, and the
other variable on the imaginary time branch.\cite{Stefanucci} 
They
do not have a specific name, however we will refer them to the right and left time mixing
Green function, respectively. The left corner $2\times 2$ matrix in the Wagner matrix
representation in Eq.~(\ref{wagn}) is the Keldysh representation of the nonequilibrium
Green functions in the Keldysh perturbation theory.\cite{Keldysh} The Matsubara Green
function is just the equilibrium Green function at temperature $1/\beta$. It couples
with the Keldysh Green function through the time mixing Green functions. If the
time mixing Green functions are neglected the Kadanoff-Baym-Wagner formalism reduces to
the Keldysh formalism. 

Within the Wagner matrix representation 
the Dyson equation for the nonequilibrium Green functions
can be written in the standard form like in the equilibrium case
\begin{equation}
\widehat{G} = \widehat{G}_0 +  \widehat{G}_0 \bullet   \widehat{\Sigma}
\bullet \widehat{G} , \label{dyson} 
\end{equation}
where $\widehat{G}_0$ is the bare Green function, and $\widehat{\Sigma}$ is the
self energy. The self energy is also written in the Wagner matrix representation
\begin{equation}
\widehat{\Sigma} =
\left(
\begin{array}{ccc}
\Sigma^{R} & \Sigma^{K} & \sqrt{2} \Sigma^{\rceil} \\
0 & \Sigma^{A} & 0 \\
0 & \sqrt{2} \Sigma^{\lceil} & \Sigma^{M} 
\end{array}
\right) . \label{wagnself}
\end{equation}
Note that in the Dyson equation (\ref{dyson}) the product symbol $\bullet$ denotes 
not only the matrix
multiplication, but also the integration over the time variables. We also omitted other
variable notations such as of momentum or spin to simplify the equation writing.
We introduce the inverse
matrix Green function $\widehat{\widetilde{G}}$ by the standard definition
\begin{equation}
\widehat{\widetilde{G}} \bullet \widehat{G} = \widehat{1} .
\label{invident}
\end{equation}
The inverse matrix Green function is also presented in the Wagner matrix representation.
One can find its elements by explicitly writing the component equations of
Eq.~(\ref{invident})
\begin{eqnarray}
\widetilde{G}^{R/A} \cdot G^{R/A} = \hat{1}, 
\label{invg1}\\
\widetilde{G}^{M} \star G^{M} = \hat{1}, \\
\widetilde{G}^{R} \cdot G^{K} + \widetilde{G}^{K} \cdot G^{A}
+ 2 \widetilde{G}^{\rceil} \star G^{\lceil} = 0, \label{invg3} \\
\widetilde{G}^{R} \cdot G^{\rceil} + \widetilde{G}^{\rceil} \star G^{M} = 0, \\ 
\widetilde{G}^{\lceil} \cdot G^{A} + \widetilde{G}^{M} \star G^{\lceil} = 0 .
\label{invg5} 
\end{eqnarray}
Here the dot and star products are the integrations over the real time and the imaginary time variables, respectively, i.e.,
\begin{eqnarray*} 
(A \cdot B)(\bar{t},\bar{t}') &=& \int_{t_0}^{t_m} dt_1 A(\bar{t},t_1) B(t_1,\bar{t}') , \\
(A \star B)(\bar{t},\bar{t}') &=& -i \int_{0}^{\beta} d\tau_1 
A(\bar{t},\tau_1) B(\tau_1,\bar{t}') .
\end{eqnarray*} 
The symbol $\hat{1}$ is just the delta function of the time variables. For real time variables it is $\delta(t-t')$,
and for imaginary time variables it is $i \delta(\tau-\tau')$. One can view
$\widetilde{G}^{R/A/M}$ as the inverse matrices of ${G}^{R/A/M}$ in 
continuous time variables. However, $\widetilde{G}^{K/\rceil/\lceil}$ are not inverses
of the corresponding Green functions. The Dyson equation (\ref{dyson}) can be rewritten
as follows
\begin{equation}
\widehat{\widetilde{G}}_0 \bullet \widehat{G} = \widehat{1} +
\widehat{\Sigma} \bullet \widehat{G} , \label{dyson1} 
\end{equation}
where $\widehat{\widetilde{G}}_0$ is the inverse matrix of $\widehat{G}_0$,
and its elements can be found from Eqs.~(\ref{invg1})-(\ref{invg5}) for the bare
Green functions.
Equation (\ref{dyson1}) can be written 
in the explicit form for the component Green functions
\begin{eqnarray}
\widetilde{G}_{0}^{R/A}  \cdot G^{R/A} &=& \hat{1} 
+  \Sigma^{R/A} \cdot G^{R/A}, \label{kb1}\\
\widetilde{G}_{0}^{M} \star G^{M} &=& \hat{1} + 
\Sigma^{M} \star G^{M}, \label{kb2} \\
\widetilde{G}_{0}^{R} \cdot G^{\rceil} &=&  \Sigma^{R} \cdot G^{\rceil} +
\big(\Sigma^{\rceil} - \widetilde{G}_0^{\rceil} \big) \star G^{M} , \label{kb3} \\
\widetilde{G}_{0}^{M} \star G^{\lceil} &=&  \Sigma^{M} \star G^{\lceil}+
\big(\Sigma^{\lceil} - \widetilde{G}_0^{\lceil} \big) \cdot G^{A} , \label{kb4}\\
\widetilde{G}_{0}^{R}  \cdot G^{K} &=&  \Sigma^{R} \cdot G^{K} + 
\big(\Sigma^{K} - \widetilde{G}_0^{K} \big) \cdot G^{A} 
\nonumber \\
&& + 2 \big(\Sigma^{\rceil} - \widetilde{G}_0^{\rceil} \big) \star G^{\lceil} .
\label{kb5}
\end{eqnarray} 
Equations (\ref{kb1})-(\ref{kb5}) are just the Kadanoff-Baym equations for the
nonequilibrium Green functions written in the Wagner representation. 
In the standard Kadanoff-Baym equations\cite{Kadanoff,Wagner} the 
inverse bare Green functions are written in the form of 
differential operators, and these differential equations also require 
additional boundary conditions.
In the Kadanoff-Baym-Wagner equations (\ref{kb1})-(\ref{kb5}) the inverse bare Green functions have
their explicit forms and they already satisfy their boundary conditions.
Instead of differential-integral equations in the Kadanoff-Baym formalism of 
the contour-ordered Green function, the Kadanoff-Baym-Wagner equations (\ref{kb1})-(\ref{kb5}) 
are just only the integral equations. 
Once 
the self energy is computable the Kadanoff-Baym-Wagner equations 
(\ref{kb1})-(\ref{kb5}) can be solved. First we solve 
Eqs.~(\ref{kb1})-(\ref{kb2}) for the retarded, advanced and Matsubara Green functions.
These equations can be solved independently. Certainly, the advanced Green function
can be quickly obtained from the retarded Green function by the relation
\begin{eqnarray}
G^{A}(t,t')= \big[G^{R}(t',t) \big]^{*} .
\end{eqnarray} 
Moreover,
the Matsubara equation (\ref{kb2}) is the equilibrium equation and we can also use the
equilibrium techniques to calculate the Matsubara Green function. 
Next we use the retarded, advanced and Matsubara Green functions as the inputs and solve
the next two equations for the time mixing Green functions. Finally, we solve the last
equation for the Keldysh Green function. The Kadanoff-Baym-Wagner equation for the Keldysh Green function (\ref{kb5}) can be rewritten as
\begin{eqnarray}
G^{K} &=& \big(\hat{1} + G^{R} \cdot \Sigma^{R} \big) \cdot G_0^{K}  \cdot 
\big(\hat{1}+ \Sigma^{A} \cdot G^{A} \big)  \nonumber \\
&& + G^{R} \cdot \Sigma^{K}  \cdot G^{A} 
 - 2 G^R \cdot \Big[ \widetilde{G}_0^{\rceil} \star G^{M}_0 \star \widetilde{G}_0^{\lceil} \nonumber \\
&& - \big(\Sigma^{\rceil} - \widetilde{G}_0^{\rceil} \big) \star G^{M}  \star \big(\Sigma^{\lceil} - \widetilde{G}_0^{\lceil} \big) 
 \Big] \cdot G^{A} .
\label{kbk}
\end{eqnarray}
Here we have used Eqs.~(\ref{invg3}), (\ref{invg5}) 
for $\widetilde{G}_0^{K}$, $\widetilde{G}_0^{\lceil}$  and Eqs.~(\ref{kb1}), 
(\ref{kb4}) for the retarded (advanced) and time mixing Green functions.
If the time mixing Green functions
are neglected, the Kadanoff-Baym-Wagner equation (\ref{kbk}) is reduced to the
Keldysh equation
\begin{eqnarray}
G^{K} 
&=& (\hat{1} + G^{R} \cdot \Sigma^{R} ) \cdot G_0^{K}  \cdot (\hat{1}+ 
\Sigma^{A} \cdot G^{A} )
\nonumber \\ 
&& + G^{R} \cdot \Sigma^{K} \cdot G^{A} . 
\end{eqnarray}
The Keldysh formalism neglects 
the contributions generated from the dynamics of the system in the imaginary time branch
of the Kadanoff-Baym contour.  
Since at the initial time $t_0$ the system is prepared in
equilibrium with full interaction, the neglected contributions are
correlations of electrons between the initial time and an advanced time. 
Indeed, if we neglect
the correlation effects of the Matsubara and the time mixing Green functions 
(i.e., $\Sigma^{M/\rceil/\lceil}=0$), 
the last
term in Eq.~(\ref{kbk}) vanishes, and we again obtain the Keldysh equation.
The neglected contributions are called initial correlations.\cite{Rammer,Fujita,Hall,Wagner}
The initial correlations distinguish between the Kadanoff-Baym-Wagner and the Keldysh
formalisms. One can notice that the equations for the retarded and advanced Green functions
are decoupled from the system of equations, 
hence the nonequilibrium density of states remains the same in both the 
Kadanoff-Baym-Wagner and the Keldysh formalisms. The initial correlations do not affect the
nonequilibrium density of states. They affect only the nonequilibrium distribution function.
Thus the initial correlations give contributions only to physical quantities which depend on 
the nonequilibrium distribution function.  

The model we will study is the FKM with external electric field turned on at $t=0$.
At the initial time $t_0$ the system is prepared in equilibrium with 
temperature $1/\beta$ and the FKM Hamiltonian 
\begin{eqnarray}
H_{\text{eq}} &= - & \sum_{i,j} J_{ij} c^{\dagger}_i c_{j} 
- \mu \sum_{i} c^{\dagger}_i c_i + E_f \sum_{i} f^{\dagger}_i f_i \nonumber \\
&& + U \sum_{i} c^{\dagger}_i c_i f^{\dagger}_i f_i , 
\label{fkm}
\end{eqnarray}
where $c^{\dagger}_i$ ($c_i$) are the creation (annihilation) operators for 
conduction electrons at site $i$, and $f^{\dagger}_i$ ($f_i$) are the creation (annihilation) operators for localized electrons at site $i$. $J_{ij}$ is
the hopping matrix of conduction electrons, and it is equal to $J$ for nearest
neighbor sites, and it is $0$ otherwise. $U$ is the strength of the interaction  
between the conduction and localized electrons. $\mu$ and $E_f$ are the chemical potentials
of the conduction and localized electrons, respectively. In this paper
we will only consider the 
half filling case. It turns out that in the half filing case $\mu=-E_f=U/2$.
At time $t=0$ a spatially uniform electric field is turned on. We choose the gauge
with vanishing of the scalar potential for the electric field. As a result the
electric field is described by a spatially uniform vector potential 
$\mathbf{A}(t)=-\theta(t) \mathbf{E} t$. The electric field couples to the conduction
electrons through the Peierls substitution for the hopping matrix
\begin{eqnarray}
J_{ij} &\rightarrow& J_{ij} 
\exp\Big[ -i e \int_{\mathbf{R}_i}^{\mathbf{R}_j} d\mathbf{r} 
\mathbf{A}(\mathbf{r},t) \Big] \nonumber \\
&=& J_{ij} 
\exp\big[ -i e \mathbf{A}(t) (\mathbf{R}_j-\mathbf{R}_i) \big] .
\label{J}
\end{eqnarray} 
By replacing the hopping matrix in Hamiltonian in Eq.~(\ref{fkm}) by Eq.~(\ref{J}) we
obtain full nonequilibrium Hamiltonian of the system. This nonequilibrium
FKM was introduced by Freericks {\em et al} in the study of 
the NEDMFT.\cite{Freericks} The considered nonequilibrium FKM differs from the
equilibrium FKM only by the bare energy spectra
\begin{eqnarray}
\varepsilon(\mathbf{k},t) = 
\varepsilon\big(\mathbf{k}-e \mathbf{A}(t)\big) =
- 2J \sum_{i=1}^{d} \cos\big(k_i - e A_i(t) \big) ,
\end{eqnarray}
where $d$ is the space dimension of the system. We will consider the case when
the electric field lies along the elementary cell diagonal
$$
\mathbf{A}(t) = A(t)(1,1,...,1) .
$$
In this case the bare energy spectra becomes
\begin{eqnarray}
\varepsilon(\mathbf{k},t) = \cos(e A(t)) \varepsilon(\mathbf{k})
+\sin(e A(t)) \bar{\varepsilon}(\mathbf{k}) ,
\end{eqnarray}
where 
\begin{eqnarray*}
\varepsilon(\mathbf{k}) &=& -2 J \sum_{i=1}^{d} \cos(k_i) , \\
\bar{\varepsilon}(\mathbf{k}) &=& -2 J \sum_{i=1}^{d} \sin(k_i) .
\end{eqnarray*}
In the limit of infinite dimensions $d \rightarrow \infty$ the bare density of states
has a double Gaussian form
\begin{eqnarray}
\rho(\varepsilon,\bar{\varepsilon}) = \rho_{0}(\varepsilon) 
\rho_{0}(\bar{\varepsilon}) , \label{gauss}
\end{eqnarray}
where $\rho_{0}(\varepsilon) = \exp(-\varepsilon^2)/\sqrt{\pi}$. Here we have used
$J^{*}=J \sqrt{d}$ as the unit of energy.

The nonequilibrium bare Green functions can be found from the equations of motion. 
The equation of motion for the retarded Green function reads
\begin{eqnarray*}
\big[ i \partial_t + \mu_0 - \varepsilon(\mathbf{k},t) \big]
G_{0}^{R}(\mathbf{k}|t,t') = \delta(t-t') ,
\end{eqnarray*}
where $\mu_0$ is the chemical potential of the noninteracting conduction electrons.
At half filling $\mu_0=0$. With the boundary condition
$G_{0}^{R}(\mathbf{k}|t,t)=-i$ we can find the bare nonequilibrium retarded Green
function
\begin{eqnarray}
G_{0}^{R}(\mathbf{k}|t,t') = - i \theta(t-t')
e^{i \mu_0 (t-t')} e^{ -i \int_{t'}^{t} dt_1 \varepsilon(\mathbf{k},t_1) } .
\label{g0ret}
\end{eqnarray} 
Similarly, one can find the bare advanced, Keldysh and Matsubara Green functions
\begin{eqnarray}
G_{0}^{A}(\mathbf{k}|t,t') &=& i \theta(t'-t)
e^{i \mu_0 (t-t')} e^{ -i \int_{t'}^{t} dt_1 \varepsilon(\mathbf{k},t_1) } , \\
G_{0}^{K}(\mathbf{k}|t,t') &=& i \big[2 f(\varepsilon(\mathbf{k})-\mu_0) -1 \big]
e^{i \mu_0 (t-t')} \nonumber \\
&& e^{ -i \int_{t'}^{t} dt_1 \varepsilon(\mathbf{k},t_1) } , \\
G_{0}^{M}(\mathbf{k}|\tau,\tau') &=& -i 
\big[\theta(\tau-\tau') - f(\varepsilon(\mathbf{k})-\mu_0) \big] \nonumber \\
&& e^{-(\varepsilon(\mathbf{k})- \mu_0 ) (\tau-\tau')} ,  
\end{eqnarray} 
where $f(\varepsilon)=1/(\exp(\beta \varepsilon)+1)$ is the Fermi-Dirac distribution function. The bare right time mixing Green function can be found from the
equation of motion
\begin{eqnarray*}
\big[ i \partial_t + \mu_0 - \varepsilon(\mathbf{k},t) \big]
G_{0}^{\rceil}(\mathbf{k}|t,\tau') = 0 ,
\end{eqnarray*}  
with the boundary condition $G_{0}^{\rceil}(\mathbf{k}|t_0,\tau') = G^{M}_0(0,\tau')$.
We obtain
\begin{eqnarray}
G_{0}^{\rceil}(\mathbf{k}|t,\tau') &=& - i 
\big[\theta(-\tau') - f(\varepsilon(\mathbf{k})-\mu_0) \big] 
e^{(\varepsilon(\mathbf{k})- \mu_0 ) \tau'} \nonumber \\
&& e^{i \mu_0 (t-t_0)} e^{ -i \int_{t_0}^{t} dt_1 \varepsilon(\mathbf{k},t_1) } 
\nonumber \\
&=& i G_{0}^{R}(\mathbf{k}| t,t_0) G_{0}^{M}(\mathbf{k}| 0,\tau') ,
\end{eqnarray} 
since $t \geq t_0$.
Similarly, the bare left time mixing Green function is
\begin{eqnarray}
G_{0}^{\lceil}(\mathbf{k}|\tau,t') &=& - i 
\big[\theta(\tau) - f(\varepsilon(\mathbf{k})-\mu_0) \big] 
e^{-(\varepsilon(\mathbf{k})- \mu_0 ) \tau} \nonumber \\
&& e^{i \mu_0 (t'-t_0)} e^{ i \int_{t_0}^{t'} dt_1 \varepsilon(\mathbf{k},t_1) } 
\nonumber \\
&=& - i  G_{0}^{M}(\mathbf{k}| \tau, 0) G_{0}^{A}(\mathbf{k}| t_0,t') .
\label{g0lft}
\end{eqnarray} 
The nonequilibrium bare Green functions clearly satisfy their boundary conditions. 
When the self energy is computable, it together with the bare Green functions
fully determine the nonequilibrium Green functions via the Kadanoff-Baym-Wagner
equations (\ref{kb1})-(\ref{kb5}). In the next sections we will solve 
the Kadanoff-Baym-Wagner equations with the self energy
calculated within the truncated and self-consistent perturbation theories as well as within
the NEDMFT.

\section{Truncated perturbation theory}
In this section we calculate the nonequilibrium Green functions and 
the electric current within the truncated perturbation theory of second order in $U$.
The perturbation calculations were previously performed within the Keldysh nonequilibrium 
perturbation theory,\cite{Turkowski} where the initial correlations are neglected. 
The purpose of this section is to find the contributions of the initial
correlations to the electric current within the truncated perturbation theory.

In the half filling case the first-order perturbation contributions to the self energy
vanish.\cite{Turkowski} The second-order self energy can be found by
expanding the contour-ordered Green function to second order in $U$. One
can find\cite{Turkowski}
\begin{eqnarray}
\Sigma^{\alpha}_2(t,t') = U^2 n_f (1-n_f) \frac{1}{N} \sum_{\mathbf{k}} 
G^{\alpha}_0(\mathbf{k}|t,t') , \label{pertse2}
\end{eqnarray}
where $\alpha=R,A,M,K,\rceil,\lceil$, and $n_f=1/2$ is the density of the localized
electrons at half filling. Within the second-order perturbation the self energy does
not depend on momentum. This feature is similar to the DMFT where the self energy is a function
of time variables only. Using the Dyson equation (\ref{dyson}) we can obtain the
 nonequilibrium Green functions up to second-order in $U$
\begin{eqnarray}
G_2^{\alpha}(\mathbf{k}|t,t') = G_0^{\alpha}(\mathbf{k}|t,t') 
+ \Delta G_2^{\alpha}(\mathbf{k}|t,t') , \label{pertg2}
\end{eqnarray}
where
\begin{eqnarray}
\Delta G_2^{R/A} &=& G_0^{R/A} \cdot \Sigma^{R/A}_2 \cdot G_0^{R/A} , \\
\Delta G_2^{M} &=& G_0^{M} \star \Sigma^{M}_2 \star G_0^{M} , \\
\Delta G_2^{\lceil} &=& G_0^{\lceil} \cdot \Sigma^{A}_2 \cdot G_0^{A}
+ G_0^{M} \star \Sigma^{\lceil}_2 \cdot G_0^{A} \nonumber \\ 
&& + G_0^{M} \star \Sigma^{M}_2 \star G_0^{\lceil} , \\
\Delta G_2^{\rceil} &=& G_0^{R} \cdot \Sigma^{R}_2 \cdot G_0^{\rceil}
+ G_0^{R} \cdot \Sigma^{\rceil}_2 \star G_0^{M} \nonumber \\ 
&& + G_0^{\rceil} \star \Sigma^{M}_2 \star G_0^{M} , \\
\Delta G_2^{K} &=& G_0^{R} \cdot \Sigma^{R}_2 \cdot G_0^{K}
+ G_0^{R} \cdot \Sigma^{K}_2 \cdot G_0^{A} \nonumber \\
&& +G_0^{K} \cdot \Sigma^{A}_2 \cdot G_0^{A} 
+ 2 G_0^{R} \cdot \Sigma^{\rceil}_2 \star G_0^{\lceil} \nonumber \\
&& + 2 G_0^{\rceil} \star \Sigma^{\lceil}_2 \cdot G_0^{A}
+ 2 G_0^{\rceil} \star \Sigma^{M}_2 \star G_0^{\lceil} .
\label{pertkel2}
\end{eqnarray}
Here, for simplicity we omitted the variable notations in the Green functions
and the self energy. In comparison to the Keldysh 
perturbation theory, the Kadanoff-Baym-Wagner perturbation expansions of the retarded and advanced
Green functions remain unchanged.\cite{Turkowski} However, the second-order perturbation expansion of the
Keldysh Green function is different. It consists of two parts. The first part is 
the first three terms in Eq.~(\ref{pertkel2}) which are also the perturbation contributions within the 
Keldysh perturbation theory,\cite{Turkowski} and the second part is 
the remaining last three terms  which are additional contributions generated from the 
initial correlations. The Keldysh perturbation theory neglects the second part.   

The electric current can be calculated by evaluating
\begin{eqnarray}
\mathbf{I}(t) = -i e \frac{1}{N} \sum_{\mathbf{k}} \mathbf{v}[\mathbf{k}-e\mathbf{A}(t)]
G^{<}(\mathbf{k}|t,t) , \label{current}
\end{eqnarray}
where $v_i(\mathbf{k})=J^{*} \sin(k_i)/\sqrt{d}$ is the velocity component, and
$G^{<}(\mathbf{k}|t,t)$ is the equal time lesser Green function, which can be calculated from
the Keldysh Green function by the relation
\begin{equation}
G^{<}(\mathbf{k}|t,t) = \frac{1}{2} \big( G^{K}(\mathbf{k}|t,t) + i \big) .
\end{equation}
When the electric field lies along the diagonal, all components of the electric current are equal,
and the magnitude of the current is
\begin{equation}
I(t) = \sqrt{d} \; \mathbf{I}_i(t) . \label{currentmagn}
\end{equation}
By inserting the second order perturbation expansions of the self energy in Eq.~(\ref{pertse2}) and of 
the Green functions in Eq.~(\ref{pertg2}) into the current formulas 
in Eqs.~(\ref{current})-(\ref{currentmagn}),
we obtain the electric current up to order $U^2$
\begin{eqnarray}
I_2(t) = I_0(t) + \Delta I_2(t)  , 
\end{eqnarray}
where $I_0(t)$ and $\Delta I_2(t)$ is the zeroth and second order contributions to the current.
The zeroth order current is
\begin{eqnarray}
I_0(t)= j_0 \int d\varepsilon \rho_0(\varepsilon) \varepsilon f(\varepsilon)
\sin(e A(t)) ,
\end{eqnarray}
where $j_0= e/\sqrt{d}$. It is the electric current in the noninteraction case. 
It exhibits the Bloch oscillations
with period $2 \pi/E$, and its amplitude is independent on time. 
In the noninteraction case the Bloch oscillations
of the current occur when the noninteracting electrons move in a lattice under a 
constant electric field. In this case the system is a perfect conductor, the periodicity
of the lattice restricts the wave vector to lie in the first Brillouin zone that leads to
the oscillations of the current. 
\begin{widetext}
After some analytical calculations we also obtain the second-order perturbation contributions 
to the current strictly in the half filling case
\begin{eqnarray}
\Delta I_2(t) &=& \Delta I_2^K(t) + \Delta I_2^{ic}(t) , \\
 \Delta I_2^{K}(t) &=& j_0 \frac{U^{2}}{4} 
 \int_{t_0}^{t} dt_1 \int_{t_0}^{t_1} dt_2 \int d\varepsilon \rho(\varepsilon) 
 \tanh \Big(\frac{\beta \varepsilon}{2}\Big) 
 \exp\Big[-\frac{1}{4} C^2(t_2,t_1) -\frac{1}{2} S^2(t_2,t_1)\Big] \nonumber \\
 && \Big[\varepsilon \cos\big(\varepsilon C(t_2,t_1)\big) \sin(e A(t)) + \frac{1}{2}
  \sin\big(\varepsilon C(t2,t1)\big) S(t_2,t_1)  \cos(e A(t))\Big] \nonumber
 \\
 &&+ j_0 \frac{U^{2}}{16} 
 \int_{t_0}^{t} dt_1 \int_{t_0}^{t} dt_2 \int d\varepsilon \rho(\varepsilon)
 \tanh \Big(\frac{\beta \varepsilon}{2}\Big) 
 \exp\Big[-\frac{1}{4} C^2(t_2,t_1) -\frac{1}{2} S^2(t_2,t_1)\Big] \nonumber \\
 && \sin\big(\varepsilon C(t_2,t_1)\big)
 \Big[C(t_2,t_1) \sin(e A(t)) - S(t_2,t_1)  \cos(e A(t))\Big] ,
 \\ 
 \Delta I_2^{ic}(t) &=& j_0 \frac{U^{2}}{4} \int_{t_0}^{t} dt_1 \int d\varepsilon \int d\varepsilon'
  \rho(\varepsilon) \rho(\varepsilon')
  \exp\Big[-\frac{1}{2} S^2(t_1,t_0) + i (\varepsilon-\varepsilon') C(t_1,t_0) \Big]  \nonumber \\
&& \frac{f(\varepsilon)-f(\varepsilon')}{\varepsilon'-\varepsilon}
   \Big[  S(t_1,t_0) \cos(e A(t)) + i (\varepsilon-\varepsilon') \sin(e A(t))\Big] \nonumber \\
&& + j_0 \frac{U^{2}}{4}
 \int d\varepsilon \int d\varepsilon'
  \rho(\varepsilon) \rho(\varepsilon') f(\varepsilon) f(-\varepsilon) \varepsilon \sin(e A(t))
  \nonumber \\
&&\bigg[  \frac{e^{\beta (\varepsilon-\varepsilon')} - \beta (\varepsilon-\varepsilon') -1}
  {(\varepsilon-\varepsilon')^2} - f(\varepsilon') 
  \frac{e^{\beta (\varepsilon-\varepsilon')}+e^{-\beta (\varepsilon-\varepsilon')}-2}
  {(\varepsilon-\varepsilon')^2} 
  \bigg] .
\end{eqnarray}
\end{widetext}

Here in order to simplify the expression, we have introduced the functions 
\begin{eqnarray*}
C(t_2,t_1) &=& \int_{t1}^{t_2} d t' \cos(e A(t')) ,\\
S(t_2,t_1) &=& \int_{t1}^{t_2} d t' \sin(e A(t')) .
\end{eqnarray*}
Like the Keldysh Green function the second-order current also consists of two 
parts, $\Delta I_2^K(t)$
and $\Delta I_2^{ic}(t)$. The first part $\Delta I_2^K(t)$ is the second-order contributions within
the Keldysh perturbation theory.\cite{Turkowski} The second part $\Delta I_2^{ic}(t)$ is the contributions
of the initial correlations. The second part is beyond the Keldysh perturbation theory.

\begin{figure}[b]
\vspace{-1cm}
\includegraphics[width=0.48\textwidth]{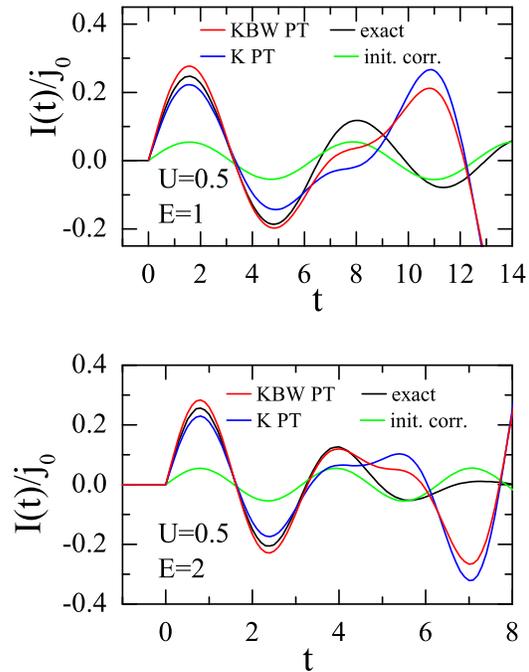}
\caption{(Color online) The time dependence of the electric current calculated within 
the Kadanoff-Baym-Wagner (KBW, red line)
and the Keldysh (K, blue line) perturbation theory (PT). The exact NEDMFT calculation result and the
initial correlation contribution to the current are presented by the black and green lines, 
respectively. The model parameters $U=0.5$, $\beta=10$, $t_0=-10$, and $E=1$ ($E=2$) for upper (lower) panel.   }
\label{fig2}
\end{figure}

\begin{figure}[t]
\vspace{-1cm}
\includegraphics[width=0.48\textwidth]{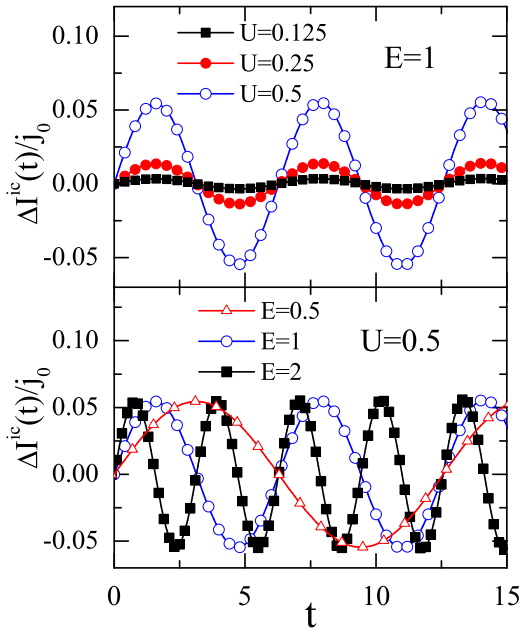}
\caption{(Color online)The second-order initial correlation contribution to the current as a function
of time for various $U$ and $E$ ($t_0=-10$, $\beta=10$).}
\label{fig3}
\end{figure}

\begin{figure}[b]
\vspace{-1cm}
\includegraphics[width=0.48\textwidth]{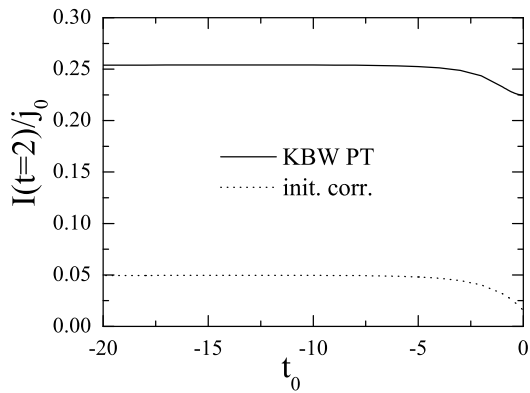}
\caption{The electric current (solid line) calculated within the Kadanoff-Baym-Wagner perturbation
theory (KBW PT) and its initial correlation part (dotted line) at 
time $t=2$ as functions of the initial
time $t_0$ for $U=0.5$, $E=1$, and $\beta=10$.}
\label{fig4}
\end{figure}

In Fig.~\ref{fig2} we plot the electric current calculated within the Kadanoff-Baym-Wagner
and the Keldysh perturbation theories up to second order in $U$. For comparison we also plot the exact
result which is obtained by performing the NEDMFT calculations (see Sec.~V). 
It shows that the current
has the Bloch oscillations with period of $2 \pi/E$ 
like the current in the noninteraction case.\cite{Freericks} 
However, the amplitude of the current varies with time.
The perturbation theories give 
reasonable results for times smaller than $\sim 2/U$. The Kadanoff-Baym-Wagner perturbation
theory overestimates the current, while the Keldysh perturbation theory  underestimates it.
 Figure~\ref{fig2} also shows that the Kadanoff-Baym-Wagner perturbation result
is closer to the exact solution than the Keldysh perturbation one at times right before the
perturbation theories are broken down. In Fig.~\ref{fig2} we also plot the initial correlation
contribution $\Delta I^{ic}_2(t)$ to the current. This part of the current also oscillates
with the same period as of the full current. In Fig.~\ref{fig3} we plot the
initial correlation part of the current for various values of $U$ and $E$. In contrast
to the full current, the amplitude of the initial correlation part does not significantly vary 
with time. 
Since the initial correlation contribution is calculated within the second-order perturbation theory, 
its amplitude is proportional to $U^2$, and almost independent on the electric field.
Usually, the Keldysh perturbation theory argues that the initial correlations vanish when
the initial time approaches to minus infinity. In Fig.~\ref{fig4} we plot the current and
its initial correlation part at a fixed time as functions of the initial time $t_0$. It
shows that both the current and its initial correlation part quickly approach to constant
values when $|t_0|$ increases. Even for $t_0=-5$ the current and its initial correlation part
already reach the constant values. The initial correlations never vanish, even when 
$t_0 \rightarrow -\infty$. Thus the Keldysh perturbation theory always neglects 
the nonvanishing initial correlations. However, within the truncated perturbation theory
both the Keldysh and the Kadanoff-Baym-Wagner formalisms only qualitatively describe the
physical properties when the perturbation theory works. The initial correlations do not
qualitatively change the perturbation results. Thus, the use of the Keldysh perturbation theory
is still convenient in the nonequilibrium study due to its simple system of equations.

\section{Self-consistent perturbation theory}
In this section we perform the self-consistent perturbation calculations for the
electric current. 
Instead of the standard perturbation calculation in Eq.~(\ref{pertse2}),
we take a self-consistent approach by dressing the bare Green functions 
 in the calculation of the self-energy, i.e.
\begin{eqnarray}
\Sigma^{\alpha}(t,t') = U^2 n_f (1-n_f) \frac{1}{N} \sum_{\mathbf{k}} 
G^{\alpha}(\mathbf{k}|t,t') . \label{selfpertse2}
\end{eqnarray}
In this approximation only the Green functions of the conduction electrons are dressed.
The Green functions of the localized electrons are kept local, thus their contributions
to the self energy of the conduction electrons are just $n_f(1-n_f)$. We solve the 
Kadanoff-Baym-Wagner equations (\ref{kb1})-(\ref{kb5}) with the self energy 
determined by Eq.~(\ref{selfpertse2}).
In order to solve these equations we adopt the discretization method which was employed
by Freericks {\em et al} in solving the NEDMFT equations.\cite{Freericks,Freericks1}
We discretize the time variables with step $\Delta t$ for real time $t$ and $\Delta \tau$ for
imaginary time $\tau$. As a result the real time domain is divided into $L$ points, 
and the imaginary time domain $\beta$ is divided  into $M$ points. 
Thus, any function of two time variables $A(\bar{t},\bar{t}')$
becomes a matrix $A_{ij}=A(\bar{t}_i,\bar{t}_j)$, where $\bar{t}_i=\bar{t}$ 
and $\bar{t}_j=\bar{t}'$. Integration over time can be
approximated by the rectangular integration rule
\begin{eqnarray*}
\int d\bar{t}_1 A(\bar{t},\bar{t}_1) B(\bar{t}_1,\bar{t}') =
\Delta \bar{t} \sum_{l} A(\bar{t}_i,\bar{t}_l) B(\bar{t}_l,\bar{t}_j) ,
\end{eqnarray*} 
where $\Delta \bar{t}=\Delta t$ for real time integration, and 
$\Delta \bar{t}=-i \Delta \tau$ for imaginary time integration. Thus 
the time integration becomes a matrix multiplication. The inverse
of the continuous matrix function
\begin{eqnarray*}
\int d\bar{t}_1 A(\bar{t},\bar{t}_1) A^{-1}(\bar{t}_1,\bar{t}') = 
\delta(\bar{t}-\bar{t}')
\end{eqnarray*} 
in the discretization approach becomes
\begin{eqnarray}
 \Delta \bar{t} \sum_{l} A(\bar{t}_i,\bar{t}_l) A^{-1}(\bar{t}_l,\bar{t}_j) = 
 \frac{\delta_{ij}}{\Delta \bar{t}} .
 \label{inv}
\end{eqnarray} 
Thus, in the discretization approach the Kadanoff-Baym-Wagner equations become the matrix equations
which can be solved numerically.
The time discretization is a numerical approach which approximately solve the 
Kadanoff-Baym-Wagner equations. It becomes exact only for 
$\Delta \bar{t} \rightarrow 0$. Nevertheless, it was shown that the discretization
approach is an efficient way to solve the nonequilibrium Green function 
equations.\cite{Freericks,Freericks1} Note that the 
Kadanoff-Baym-Wagner equations (\ref{kb1})-(\ref{kb5}) differs from the
contour-ordered Green function equation.\cite{Freericks}
Numerically, here we have to solve the equations of matrices with size 
$L\times L$, $L \times M$, and $M\times M$, instead of matrices of size 
$(2 L+M)\times (2L+M)$ in the contour-ordered Green function equation.
It reduces the matrix size and computation time. However, here we have to solve 
five equations with additional matrix multiplications. The inverse bare Green 
functions $\widetilde{G}^{\alpha}_0$ are calculated from Eqs.~(\ref{invg1})-(\ref{invg5})
with the inputs of the bare Green functions in Eqs.~(\ref{g0ret})-(\ref{g0lft}). Within
the discretization accuracy, these inverse bare Green functions are calculated exactly.
They satisfy the  boundary conditions. For instance, the Matsubara Green function
has the antiperiodic property in the time variable, or the Keldysh Green function satisfies
$G^{K}_{0}(t_0,t_0)=2 G^{M}_{0}(0,0^+)-i$. In the
contour-ordered Green function approach, the inverse bare Green function contains a time
differential operator and it is also approximately discretized. 
In the present approach the inverse bare retarded and advanced Green functions are numerically  
calculated from their bare functions by the discretization inverse relation in Eq.~(\ref{inv}). 
The inverse Matsubara Green function in the discretization
form can be analytically obtained 
\begin{widetext}
\begin{eqnarray}
\widetilde{G}^{M}_{0}(\mathbf{k}) = -\frac{i}{\Delta \tau^2}
\left( \begin{array}{ccccccc}
1 & 0 & 0 &  & \cdots &  & e^{-\varepsilon(\mathbf{k}) \Delta \tau} \\
- e^{-\varepsilon(\mathbf{k}) \Delta \tau} & 1 & 0 &  & \cdots & & 0 \\
0 & - e^{-\varepsilon(\mathbf{k}) \Delta \tau} & 1 &  0 & & & \\
\vdots & \vdots & \vdots &   &  &  & \vdots \\
0 & \cdots & \cdots & & \cdots& - e^{-\varepsilon(\mathbf{k}) \Delta \tau} & 1 
\end{array}
\right) .
\label{gm0}
\end{eqnarray}
Here we have taken into account $\mu_0=0$ at half filling.
This inverse bare Matsubara Green function corresponds to the bare Matsubara Green function
with fixed diagonal elements $G^{M}_0(\tau,\tau)=-i(1-f(\varepsilon(\mathbf{k})))$.
It is suitable for calculating the left time mixing Green function because of the boundary
condition $G_0^{\lceil}(\tau,t_0)=G_0^{M}(\tau,0)$ for $\tau \geq 0$. The right time mixing Green
function has the boundary condition $G_0^{\rceil}(t_0,\tau')=G_0^{M}(0,\tau')$ for
$\tau' \geq 0$, and the Matsubara Green function suitable for its calculations has the  
diagonal elements $G^{M}_0(\tau,\tau)=i f(\varepsilon(\mathbf{k}))$. The corresponding
inverse bare Matsubara Green function has the matrix form
\begin{eqnarray}
\widetilde{G}^{M}_{0}(\mathbf{k}) = \frac{i}{\Delta \tau^2}
\left( \begin{array}{ccccccc}
1 &  -e^{\varepsilon(\mathbf{k}) \Delta \tau} & 0 &  & \cdots &  & 0 \\
0 & 1 & -e^{\varepsilon(\mathbf{k}) \Delta \tau} & 0 & \cdots & & 0 \\
0 & 0 & 1 &  -e^{\varepsilon(\mathbf{k}) \Delta \tau} & & & \vdots \\
\vdots & \vdots & \vdots &   &  &  & -e^{\varepsilon(\mathbf{k})\Delta \tau} \\
e^{\varepsilon(\mathbf{k}) \Delta \tau} & \cdots & \cdots & & \cdots& 0 & 1 
\end{array}
\right) . \label{gm01}
\end{eqnarray}  
\end{widetext}
For a definiteness we also use this Matsubara Green function for calculating 
the Keldysh Green function.
The Green functions $\widetilde{G}^{K/\rceil/\lceil}_0$ can be also analytically obtained.
From Eqs.~(\ref{invg1})-(\ref{invg5}) for the bare Green functions in Eqs.~(\ref{g0ret})-(\ref{g0lft})
one can show that
\begin{eqnarray}
\widetilde{G}_{0}^{\rceil}(t,\tau') &=&   \delta(t-t_0) \delta(\tau') , \\
\widetilde{G}_{0}^{\lceil}(\tau,t') &=& - \delta(\tau) \delta(t'-t_0) , \\
\widetilde{G}_{0}^{K}(t,t') &=& i \delta(t-t_0) \delta(t'-t_0) . 
\end{eqnarray}
In the Keldysh formalism, when the time mixing Green functions are neglected,
the Green function $\widetilde{G}^{K}_0$ is little changed. One can obtain
\begin{eqnarray}
\widetilde{G}_{0}^{K}(t,t') &=& - i \big[ 2 f(\varepsilon(\mathbf{k}))-1\big]
\delta(t-t_0) \delta(t'-t_0) . 
\end{eqnarray}

\begin{figure}[t]
\vspace{-1cm}
\includegraphics[width=0.4\textwidth]{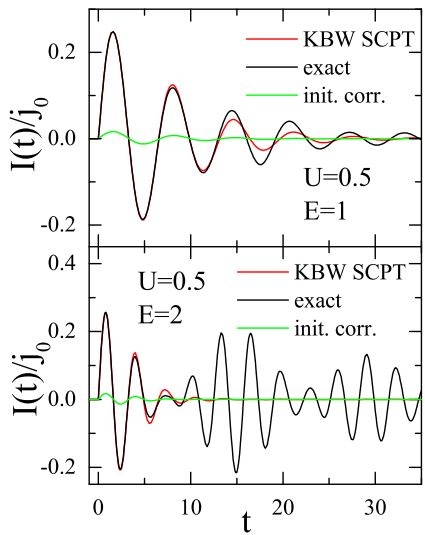}
\caption{(Color online) The time dependence of the electric current
calculated within the Kadanoff-Baym-Wagner self-consistent perturbation theory 
(KBW SCPT, red line). The exact NEDMFT calculation result and the
initial correlation contribution to the current are presented by the black and green lines, 
respectively. The current was already scaled with a quadratic extrapolation
($\Delta t=0.1$, $0.065$, $0.05$, and $\Delta \tau=0.1$).  
The model parameters $U=0.5$, $t_0=-5$, $\beta=10$, $E=1$ ($E=2$) for upper (lower) panel. }
\label{fig5a}
\end{figure}

\begin{figure}[t]
\vspace{-1cm}
\includegraphics[width=0.48\textwidth]{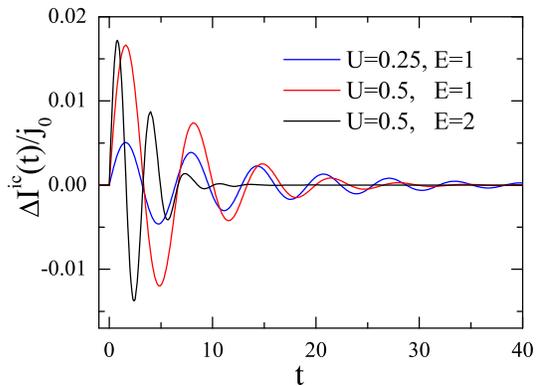}
\caption{(Color online) 
The time dependence of the initial correlation
contribution to the current
calculated within the Kadanoff-Baym-Wagner self-consistent perturbation theory 
for various $U$ and $E$. The current was already scaled with a quadratic extrapolation
($\Delta t=0.1$, $0.065$, $0.05$, and $\Delta \tau=0.1$).  
The other model parameters $t_0=-5$, $\beta=10$.}
\label{fig5b}
\end{figure}

In order to solve the Kadanoff-Baym-Wagner equations (\ref{kb1})-(\ref{kb5}), 
first we solve Eqs.~(\ref{kb1}) and (\ref{kb2}) for the retarded (advanced) and 
Matsubara Green functions, and then we find the time mixing Green functions from
Eqs.~(\ref{kb3})-(\ref{kb4}). Finally, the Keldysh Green function is calculated
from Eq.~(\ref{kb5}). We use iterations to solve each equation. When the nonequilibrium
Green functions are obtained, we can compute the electric current by Eq.~(\ref{current}). 
The momentum summation in Eq.~(\ref{selfpertse2}) or (\ref{current}) 
indeed is the integration with
the double Gaussian density of states in Eq.~(\ref{gauss}), and we use a 
Gaussian quadrature
to calculate it. Typically, we use $51$ points for the Gaussian quadrature. In the next
section we will discuss this type of integrations in a more detail. 
The calculated current converges with $\Delta t$ well. We can obtain reliable
results at $\Delta t \rightarrow 0$ by using a Lagrange interpolation formula.
Typically, we use a quadratic interpolation to obtain the current in the
continuous limit.
In Fig.~\ref{fig5a}
we plot the electric current obtained within the Kadanoff-Baym-Wagner self-consistent
perturbation theory. For comparison we also plot the exact NEDMFT calculation result (see also the next
section). Figure \ref{fig5a} shows that the self-consistent perturbation theory gives
very good results for time smaller than $2/U$. In comparison with the truncated perturbation
theory, the self-consistent perturbation theory gives reasonable results in a 
wide range of the time variable. The current obtained within the 
self-consistent perturbation theory also oscillates with time, and is damped to zero value. 
Even for large electric fields 
(for instance, $E=1$)
the time damping of the current is still observed in the self-consistent perturbation results 
like the exact solution. However, for
larger electric fields (for instance, $E=2$), 
the self-consistent perturbation theory cannot
reproduce the beat behavior of the current, as shown in the lower panel 
of Fig.~\ref{fig5a}. It shows that the self-consistent perturbation theory may
produce artifacts, especially for nonequilibrium  steady state. However, this happens only
for very strong electric fields. 
We define the initial correlation contribution to the current
by the difference of the currents calculated within the Kadanoff-Baym-Wagner and
the Keldysh self-consistent perturbation theories. In contrast to
the results of the truncated perturbation theory, the initial correlation part of
the current is damped with time, and its amplitude is significantly 
smaller. We plot the initial
correlation contribution to the current for various values of $U$ and $E$ in 
Fig.~\ref{fig5b}. It shows that the
amplitude of the initial correlation part is not scaled with $U^2$.
For long time limit the initial correlation contribution to the current vanishes.
In this case the Keldysh and the Kadanoff-Baym-Wagner formalisms give the same
steady state. However, it may be an artifact, especially for very strong electric fields when
the exact current exhibits the beat behavior. This also indicates that the self-consistent
perturbation theory may not work well for very strong electric fields.
Nevertheless,  within the self-consistent perturbation theory 
the Keldysh formalism qualitatively
gives almost the same results as of the Kadanoff-Baym-Wagner formalism. The initial correlations
do not qualitatively change the perturbation results.

\section{Nonequilibrium dynamical mean-field theory}
In this section we present the NEDMFT through the Kadanoff-Baym-Wagner representation.
The NEDMFT was proposed by Freericks {\em et al} and it is based on the same idea of the DMFT in
equilibrium.\cite{Freericks} 
The NEDMFT has the same principle features of the equilibrium DMFT. It becomes
exact in the infinite dimension limit.
In infinite
dimensions the self energy is purely local in space. It can be determined
by mapping the lattice problem into an effective problem of a single site 
embedded in a self-consistent effective 
medium. The effective medium can be represented by a Green function 
$\widehat{\mathcal{G}}$ which is determined by the Dyson equation 
\begin{eqnarray}
\widehat{G} = \widehat{\mathcal{G}}  + \widehat{\mathcal{G}} \bullet 
\widehat{\Sigma} \bullet \widehat{G}, 
\end{eqnarray}
where  $\widehat{G}=\sum_{\mathbf{k}} \widehat{G}(\mathbf{k})/N$. From this
equation we can find the components of the effective medium Green function in 
the Kadanoff-Baym-Wagner representation like Eqs.~(\ref{kb1})-(\ref{kb5}). We obtain
\begin{eqnarray}
\widetilde{\mathcal{G}}^{R/A}  &=& \widetilde{G}^{R/A} 
+  \Sigma^{R/A} , \label{dmftkb1}\\
\widetilde{\mathcal{G}}^{M}  &=& \widetilde{G}^{M} + 
\Sigma^{M}  , \label{dmftkb2} \\
\widetilde{\mathcal{G}}^{\rceil} &=& \Sigma^{\rceil} -
\widetilde{G}^{R} \cdot G^{\rceil} \star \widetilde{G}^{M} , 
\label{dmftkb3} \\
\widetilde{\mathcal{G}}^{\lceil} &=& \Sigma^{\lceil} -
\widetilde{G}^{M} \star G^{\lceil} \cdot \widetilde{G}^{A} , 
\label{dmftkb4}\\
\widetilde{\mathcal{G}}^{K} &=& \Sigma^{K} -
\widetilde{G}^{R}  \cdot G^{K} \cdot \widetilde{G}^{A} 
\nonumber \\
&& + 2 \big(\Sigma^{\rceil} - \widetilde{\mathcal{G}}^{\rceil} \big) \star G^{\lceil}
\cdot \widetilde{G}^{A} ,
\label{dmftkb5}
\end{eqnarray} 
where $\widehat{\widetilde{\mathcal{G}}}$ and $\widehat{\widetilde{G}}$  
are the inverse matrices of $\widehat{\mathcal{G}}$ and $\widehat{G}$, respectively. 
Equations (\ref{dmftkb1})-(\ref{dmftkb5}) are the Kadanoff-Baym-Wagner equations
for determining the effective medium Green function $\widehat{\widetilde{\mathcal{G}}}$. 
Once the effective medium Green function $\widehat{\widetilde{\mathcal{G}}}$ is determined
we can compute the single-site Green function. In the homogeneous phase
we obtain\cite{Freericks}
\begin{eqnarray}
\widehat{G}_{imp} = (1-n_f) \widehat{Q}_0 + n_f \widehat{Q}_1 ,
\end{eqnarray}
where $n_f$ is the localized electron density, and $\widehat{Q}_l$ 
with $l=0,1$
satisfy the following equation
\begin{eqnarray}
\big[ \widehat{\widetilde{\mathcal{G}}} + \Delta \mu - l U \big] \bullet
\widehat{Q}_l &=& \widehat{1} . \label{eqq} 
\end{eqnarray}
Here $\Delta \mu=\mu -\mu_0$. One can find explicitly the 
components of $\widehat{Q}_l$ by using the inverse equations 
(\ref{invg1})-(\ref{invg5})
\begin{eqnarray}
Q^{R/A}_l &=& 
\big[ \widetilde{\mathcal{G}}^{R/A} + \Delta \widetilde{g}_{l}^{R/A} \big]^{-1} , \\ 
Q^{M}_l &=& 
\big[ \widetilde{\mathcal{G}}^{M} + \Delta \widetilde{g}_{l}^{M} \big]^{-1} , \\
Q^{\rceil}_l &=& - Q^{R}_l \cdot
\widetilde{\mathcal{G}}^{\rceil} \star Q^{M}_l , \\
Q^{\lceil}_l &=& - Q^{M}_l \star 
\widetilde{\mathcal{G}}^{\lceil} \cdot Q^{A}_l , \\
Q^{K}_l
&=& - Q^{R}_l \cdot \widetilde{\mathcal{G}}^{K} \cdot Q^{A}_l \nonumber  \\
&& + 2 Q^{R}_l \cdot 
\widetilde{\mathcal{G}}^{\rceil} \star Q_l^M \star \widetilde{\mathcal{G}}^{\lceil}
\cdot Q^{A}_l
\end{eqnarray}
where 
$\Delta \widetilde{g}_l^{\alpha} = 
\widetilde{g}^{\alpha}(\mu_0+\Delta \mu - l U)- \widetilde{g}^{\alpha}(\mu_0)$
with $\alpha=R,A,M$, and $\widetilde{g}^{\alpha}(x)$ is the inverse of the bare
Green function $g^{\alpha}(x)$ of a pure noninteracting single site 
with zero energy level and the chemical potential $x$. Note that in the numerical calculations
when we make the discretization of the time variable, 
the quantity $\Delta \mu - l U$ does not lie in the diagonal
of the matrices of the inverse retarded (advanced) or Matsubara Green functions.
It lies in the first subdiagonal of the matrices like 
in Eqs.~(\ref{gm0}), (\ref{gm01}). In the such way we can compute the single-site
Green function $\widehat{G}_{imp}$. However, it is applicable only for 
nonequilibrium FKM. For other models such as the Hubbard model one may adopt
different techniques to solve the effective single-site problem.

The self-consistent condition requires that
\begin{eqnarray}
\widehat{G}_{imp} = \widehat{G} .
\end{eqnarray}
With this self-consistent condition when the effective single-site problem is solved
we can again compute the self energy from the Kadanoff-Baym-Wagner equations
(\ref{dmftkb1})-(\ref{dmftkb5}). When the self energy is obtained the full lattice Green
functions are calculated from the Kadanoff-Baym-Wagner equations (\ref{kb1})-(\ref{kb5}).
Thus we obtain a closed system of equations for the nonequilibrium Green functions in the
NEDMFT. Like the previous section, first we solve the set of equations of the retarded
(advanced) and Matsubara Green functions. Then use the obtained Green functions to solve
the set of equations of the time mixing Green functions. Finally, we compute
the Keldysh Green function from its set of equations. We use iterations for finding
each Green function. Numerically, we employ the discretization method 
which was described in Sec.~IV to solve the Kadanoff-Baym-Wagner NEDMFT equations.
In equilibrium the FKM describes a metal-insulator transition for the homogeneous
phase.\cite{Kennedy,Lemanski,Falicov} The Coulomb interaction is divided into two ranges. For weak
interactions ($U < \sqrt{2}$) the system is metallic, and for strong ones 
($U > \sqrt{2}$) the system is insulator. We will study the two cases separately.

\begin{figure}[b]
\vspace{-1cm}
\includegraphics[width=0.5\textwidth]{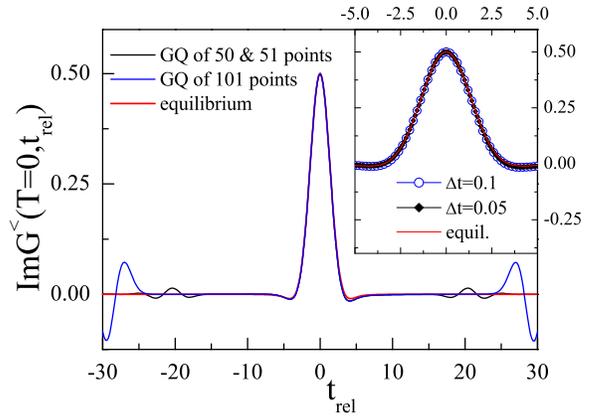}
\caption{(Color online) The imaginary part of the equilibrium lesser Green function 
$\text{Im} G^{<}(T=0,t_{rel})$ calculated
within the Kadanoff-Baym-Wagner NEDMFT by using two Gaussian quadratures (GQ) of $50$ and $51$ points 
(black line) and by one Gaussian quadrature of $101$ points (blue line) 
($\Delta t=0.1$, $\Delta \tau=0.1$, $t_0=-15$) 
in comparison with the result calculated by solving the equilibrium
DMFT equations in frequency (red line). The inset focuses the imaginary part of the equilibrium 
lesser Green function $G^{<}(T=0,t_{rel})$ obtained by performing the two Gaussian quadratures 
of $50$ and $51$ points
in a small range of $t_{rel}$ for different $\Delta t$  ($\Delta \tau=0.1$, $t_0=-15$).  
The model parameters $U=0.5$, $\beta=10$.}
\label{fig7}
\end{figure}

As a benchmark we apply the Kadanoff-Baym-Wagner NEDMFT to the equilibrium
FKM at half filling. The DMFT results of the FKM at equilibrium can be also obtained by
solving the DMFT equations in frequency.\cite{Brandt} As noted in Sec.~IV, the summation
over momentum is replaced by integration with the double Gaussian density of states 
in Eq.~(\ref{gauss}), and we use a Gaussian quadrature to perform the calculation.
Freericks {\em et al} noticed that averaging the results of two Gaussian quadratures
with $n$ and $n+1$ points works better than choosing $(2n+1)$ points for the 
quadrature.\cite{Freericks1}
For the equilibrium case we adopt this trick. 
In Fig.~\ref{fig7} we plot the lesser Green function 
$G^{<}(T,t_{rel})=\sum_{\mathbf{k}} G^{<}(\mathbf{k}|T,t_{rel})/N$
calculated within the
Kadanoff-Baym-Wagner NEDMFT in comparison with the one obtained by solving the equilibrium  DMFT
equations in frequency. Here we have converted the results from the time variables $t$ and $t'$
to Wigner average $T=(t+t')/2$ and relative $t_{rel}=t-t'$ time variables.
The NEDMFT calculations are performed with two Gaussian
quadratures with $50$ and $51$ points. We also plot the NEDMFT result which is obtained
by performing only one Gaussian quadrature with $101$ points. The results
plotted in Fig.~\ref{fig7} confirm the notice of Freericks {\em et al}. Indeed, 
for a range of small $t_{rel}$
the lesser Green function calculated within the NEDMFT fits perfectly with the one
obtained within the DMFT. However, for $t_{rel}$ nearby the time cutoffs, the NEDMFT
results exhibit spurious features of a nodal form due to finite size effects of the numerical procedures. 
These spurious features are greatly reduced by employing the trick of
two Gaussian quadratures. However, as we will see later, the spurious features
do not appear in the nonequilibrium case where the electric field is finite. 
The lesser Green function obtained within the NEDMFT fulfils the sum rule very well,
as shown in the inset of Fig~\ref{fig7}. Indeed, 
$\text{Im} G^{<}(T,t_{rel}=0) \approx 0.5$ in the weak interaction case. The sum rule of
higher-order moments of the lesser Green function is also fulfilled because the
lesser Green function fits perfectly with the DMFT one nearby $t_{rel}=0$. However,
if one numerically calculates the sum rule of higher-order moments, a deviation from
the exact value may appear due to numerical derivative calculations from discretized
points.\cite{Turkowski1} The possible deviation of the sum rule of higher order moment
does not necessarily
mean an inaccuracy of the Green function. It relates to the finite value of $\Delta t$ 
which may be not small enough  
for performing numerical derivative calculations.

\begin{figure}[b]
\vspace{-1cm}
\includegraphics[width=0.5\textwidth]{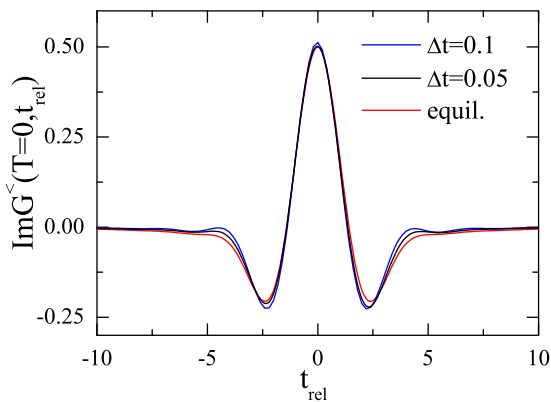}
\caption{(Color online) The imaginary part of the 
equilibrium lesser Green function $\text{Im} G^{<}(T=0,t_{rel})$ calculated
within the Kadanoff-Baym-Wagner NEDMFT for different $\Delta t$ ($\Delta \tau=0.1$, $t_0=-15$). The
two 
Gaussian quadratures with $50$ and $51$ points are performed. 
The equilibrium DMFT calculation result is also 
presented (red line).   
The model parameters $U=2$, $\beta=10$.}
\label{fig8}
\end{figure}

In Fig.~\ref{fig8} we plot the imaginary part of the 
equilibrium lesser Green function calculated within the
Kadanoff-Baym-Wagner NEDMFT and the equilibrium DMFT for $U=2$. This value of $U$
corresponds to the insulator phase. It shows that for strong interactions
the NEDMFT results fit well with the ones of the equilibrium DMFT for
small $t_{rel}$. The spectral sum rule of the lesser Green function is well fulfilled.
Indeed, for $U=2$ $\text{Im} G^{<}(T=0,t_{rel}=0)$ is equal to $0.5123$ for $\Delta t=0.1$, 
and is equal to
$0.5021$ for $\Delta t=0.05$ in comparison with the exact value $0.5$.
Around the minima in the curve of the imaginary part of the lesser Green function small deviations 
appear. The deviations can be reduced by decreasing $\Delta t$. For strong interactions
the numerical results  slightly deviate from the equilibrium values due to the finite
discretization of the time variables.  
Nevertheless, the NEDMFT calculations for the equilibrium case for both 
weak and strong interactions
show the numerical techniques employed here are accurate and controllable.

\begin{figure}[b]
\vspace{-1cm}
\includegraphics[width=0.5\textwidth]{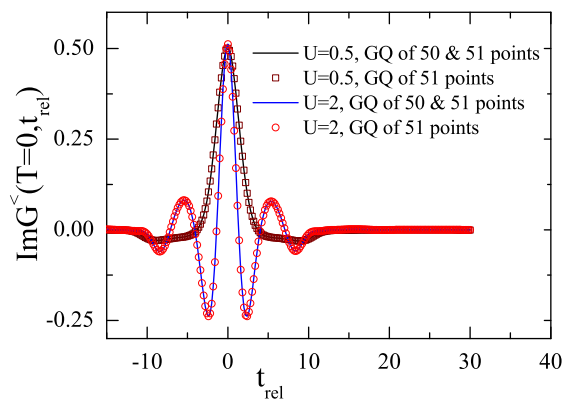}
\caption{(Color online) The imaginary part of the nonequilibrium lesser Green 
function $\text{Im} G^{<}(T=0,t_{rel})$ calculated
within the Kadanoff-Baym-Wagner NEDMFT by using two Gaussian quadratures of $50$ and $51$ 
points (solid lines), and by using one Gaussian quadrature of $51$ points (symbols) for weak ($U=0.5$) 
and strong ($U=2$) interactions ($\Delta t=0.1$, $\Delta \tau=0.1$, $t_0=-15$, $\beta=10$, $E=1$).}
\label{fig9}
\end{figure}

In the nonequilibrium case, when the electric field is finite, we notice that 
the use of two Gaussian quadratures for
the integration with the double Gaussian density of states 
gives almost the same result as the use of one Gaussian quadrature.
In Fig.~\ref{fig9} we plot the imaginary part of the lesser Green function calculated
by using two Gaussian quadratures of $50$ and $51$ points in comparison with the one
calculated by using one Gaussian 
quadrature of $51$ points for both  weak and 
strong interactions. It shows that the results of both quadrature methods are almost identical. 
In contrast
to the equilibrium case, in the nonequilibrium case there are not spurious features nearby
the time cutoffs. We have also checked the results with more points for the
Gaussian quadrature (in particular, with $n=101$), and with finer $\Delta \tau$ 
(in particular, with $\Delta \tau=0.05$). It turns out that the numerical results
are mostly sensitive to the real time discretization. In the following for numerical
 calculations we use the single Gaussian quadrature with $51$ points and make the
 integrations over the imaginary time with $\Delta \tau=0.1$ for $\beta=10$.

\begin{figure}[t]
\vspace{-1cm}
\includegraphics[width=0.5\textwidth]{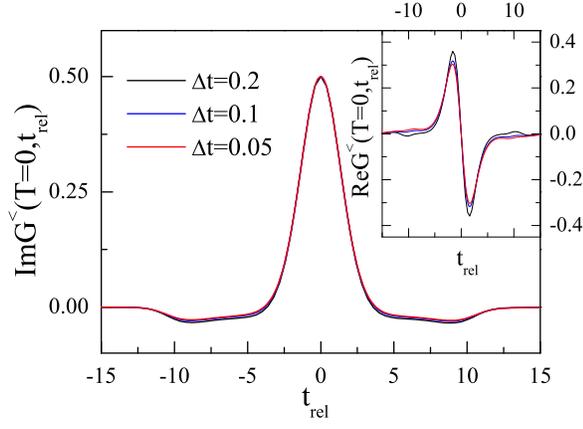}
\caption{(Color online) The imaginary part of the lesser Green function 
$G^{<}(T=0,t_{rel})$ calculated
within the Kadanoff-Baym-Wagner NEDMFT in
the weak interaction case
for various $\Delta t$. The inset plots the real part of
the lesser Green function. The model parameters $U=0.5$, $E=1$,
$\beta=10$, $t_0=-15$. $n=51$ is used for the Gaussian quadrature and 
$\Delta \tau=0.1$.}
\label{fig10}
\end{figure}
 
\begin{table}[b]
\caption{The spectral sum rule of the nonequilibrium lesser Green function at the
average time $T=0$ calculated within the
Kadanoff-Baym-Wagner NEDMFT for weak ($U=0.5$) and strong ($U=2$) interaction ($E=1$).
Various values of $\Delta t$ are used ($\Delta \tau=0.1$, $t_0=-15$, $\beta=10$).}
\begin{tabular}{c c c c c}
$U=0.5$   &  & & &  \\
\hline
\hline
$\Delta t$ & $0.2$ & $0.1$  & $0.05$ & exact  \\
sum rule & $0.5007$ & $0.5005$ & $0.5004$ & $0.5$ \\ 
\hline
\hline \\
$U=2$ & & & & \\
\hline
\hline
$\Delta t$ & $0.1$ & $0.05$ & $0.025$ & exact \\
 sum rule & $0.5123$ & $0.5021$ & $0.4983$ & $0.5$ \\ 
\hline
\hline
\end{tabular}
\end{table}

\begin{figure}[t]
\vspace{-1cm}
\includegraphics[width=0.5\textwidth]{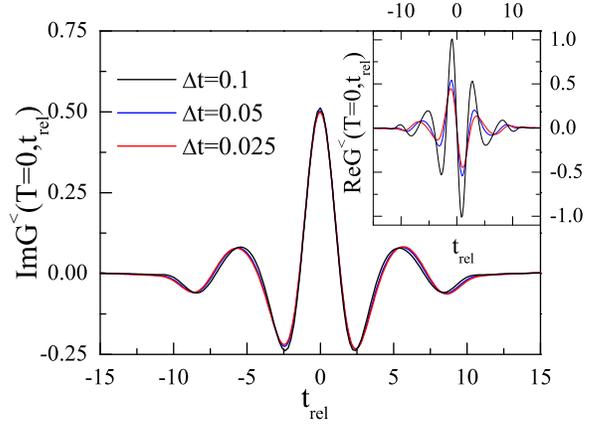}
\caption{(Color online) The imaginary part of the lesser Green function 
$G^{<}(T=0,t_{rel})$ calculated
within the Kadanoff-Baym-Wagner NEDMFT in
the strong interaction case
for various $\Delta t$. The inset plots the real part of
the lesser Green function. The model parameters $U=2$, $E=1$,
$\beta=10$, $t_0=-15$. $n=51$ for the Gaussian quadrature and 
$\Delta \tau=0.1$.}
\label{fig11}
\end{figure}

In Fig.~\ref{fig10} we present the lesser Green function in the weak
interaction case for various real time discretizations  
$\Delta t$ and a fixed $t_0$. It shows that the imaginary part of
the lesser Green function quickly converges with $\Delta t$. It also
indicates that the spectral sum rule of the lesser Green function is fulfilled
well. Indeed, in Table~I we list the value of the spectral sum rule of the
lesser Green function for various values of $\Delta t$. In the 
Kadanoff-Baym-Wagner formalism the spectral sum rule which is  numerically calculated
is fulfilled better than in the formalism of the contour-ordered Green function.\cite{Turkowski1} 
The spectral sum rule of the retarded and advanced
Green functions is fulfilled well too. The real part
of the lesser Green function converges with $\Delta t$ is less quickly. However,
it also converges well for small $\Delta t$. For weak interactions the numerical
calculations solving the Kadanoff-Baym-Wagner NEDMFT equations work very well. For other
physical quantities such as the electric current and the double occupation 
their convergences with $\Delta t$ are also good. In particular, we can obtain 
converged results at the limit $\Delta t \rightarrow 0$ by using a Lagrange
interpolation formula.    

\begin{figure}[b]
\vspace{-1cm}
\includegraphics[width=0.45\textwidth]{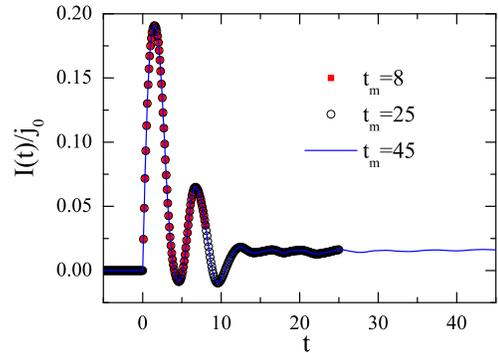}
\caption{(Color online) The time dependence of the electric current $I(t)/j_0$ calculated
within the Kadanoff-Baym-Wagner NEDMFT for various time cutoffs $t_m$
with fixed $\Delta=0.1$ and $t_0=-5$ 
($U=2$, $E=1$ $\Delta \tau=0.1$, $\beta=10$).}
\label{fig12}
\end{figure}

In Fig.~\ref{fig11} we plot the lesser Green function in the strong interaction
case for various real time discretizations  $\Delta t$ and a fixed $t_0$.
The imaginary part of the lesser Green function converges with $\Delta t$ well. However,
its spectral sum rule slightly deviates from the exact value, as presented in Table~I.
The real part of the lesser Green function converges with $\Delta t$ not so fast as
in the weak interaction case. In general, for strong interactions 
the extrapolations of the numerical
results  of the Kadanoff-Baym-Wagner NEDMFT in the limit $\Delta t \rightarrow 0$
require a more effort. Often in order to obtain reliable data we have to 
carry the numerical calculations with $\Delta t$ smaller than the ones in the weak 
interaction case.

One can notice that the numerical results of the Kadanoff-Baym-Wagner NEDMFT
are independent on the maximal time $t_m$ when $\Delta t$ and $t_0$ are fixed.
Indeed, we plot the electric current $I(t)$ obtained from the
Kadanoff-Baym-Wagner NEDMFT for various $t_m$ in Fig.~\ref{fig12}. 
For larger $t_m$ the time window is larger and we can observe the behaviors
of the system at a more long time. But for larger $t_m$ the numerical calculations
are also more expensive in time. We have to compromise the computation time and the
need of the time window width. 

\begin{figure}[t]
\vspace{-1cm}
\includegraphics[width=0.45\textwidth]{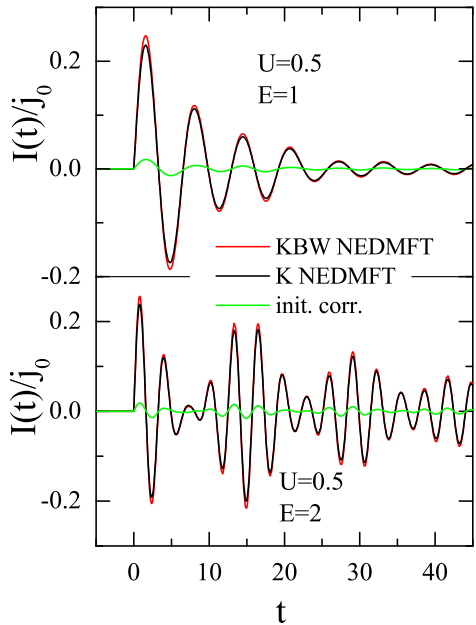}
\vspace{-0.5cm}
\caption{(Color online) The time dependence of the electric current $I(t)/j_0$ calculated
within the Kadanoff-Baym-Wagner (KBW) NEDMFT (red line) in the weak interaction case for different
electric fields. For comparison the result obtained from the Keldysh (K) NEDMFT (black line),
and the contribution of the initial correlations to the current (green line) are also plotted. 
The data are already scaled with a quadratic extrapolation 
($\Delta t=0.1$, $0.065$, and $0.05$). 
The model parameters $U=0.5$, $\beta=10$, $\Delta \tau=0.1$, and $E=1$ ($E=2$) for
upper (lower) panel.}
\label{fig13}
\end{figure}

\begin{figure}[t]
\vspace{-1cm}
\includegraphics[width=0.45\textwidth]{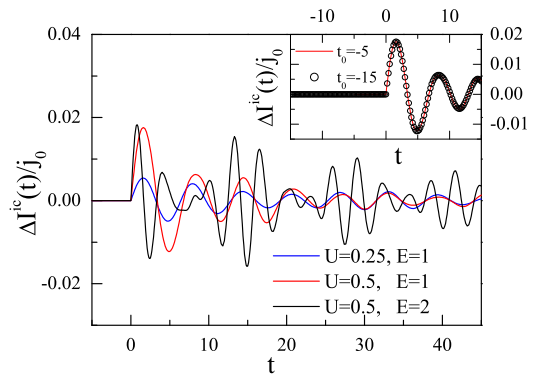}
\vspace{-0.5cm}
\caption{(Color online) The time dependence of the contribution of the initial
correlations to the electric current $\Delta I^{ic}(t)/j_0$ for various
$E$ and $U$ in the weak interaction case. The results are already scaled by
a quadratic extrapolation with $\Delta t=0.1$, $0.065$, and $0.5$
($t_0=-5$, $\Delta \tau=0.1$, $\beta=10$). The inset plots the scaled
contribution of the initial
correlations $\Delta I^{ic}(t)/j_0$ for different initial times
$t_0$ ($U=0.5$, $E=1$, $\Delta \tau=0.1$, $\beta=10$).}
\label{fig14}
\end{figure}

In Fig.~\ref{fig13} we present the electric current calculated within both the
Kadanoff-Baym-Wagner and the Keldysh NEDMFT in the weak interaction case.
In the Keldysh NEDMFT the initial correlations are neglected.
The contribution of the initial correlations to the current is defined as the difference
of the currents calculated within the Kadanoff-Baym-Wagner and the Keldysh NEDMFT.
Certainly, the current has been previously calculated within the NEDMFT of the
contour-ordered Green function.\cite{Freericks} We find 
after extrapolating to $\Delta t \rightarrow 0$ the results
of the Kadanoff-Baym-Wagner formalism agree well with the ones obtained
within the contour-ordered Green function NEDMFT. 
This indicates the equivalence of the Kadanoff-Baym-Wagner and 
the contour-ordered Green function formalisms, as expected.
However, the Kadanoff-Baym-Wagner formalism represents the contour-ordered
Green function in the matrix form, the elements of which are the physical Green functions. 
It is also similar to the Keldysh formalism. 
The spectral sum rule obtained within the Kadanoff-Baym-Wagner
NEDMFT is fulfilled very well.
The electric current displays the Bloch
oscillations, as noticed by Freericks {\em et al}.\cite{Freericks}
For small and large electric fields (for instance, $E=1$) the current is monotonously damped to zero value.
However, when the electric
field increases further (for instance, $E=2$) the current develops beats. 
As shown in Fig.~\ref{fig13}, in the weak interaction case 
the Keldysh and the Kadanoff-Baym-Wagner NEDMFT qualitatively give the same current.  
The contribution of the initial correlations to the current also oscillates with
time in the same way as of the full current. When the current displays beats the initial
correlation contribution displays beats too, as shown in Figs.~\ref{fig13} and
\ref{fig14}.  In the weak interaction case 
the initial correlation contribution to the current is small in comparison with the full current. However,
it never vanishes except for $t<0$ when the electric field is absent and the 
current vanishes too. In the inset of Fig.~\ref{fig14} we plot the initial correlation
contribution to the current for different initial times $t_0$. It shows that the results are
independent on the initial time if it is far enough from $t=0$. The initial correlation
contribution seems to be finite even when the initial time is in the remote past.
However, the initial correlations do not qualitatively change the current properties in
the weak interaction case. 

\begin{figure}[t]
\vspace{-1cm}
\includegraphics[width=0.5\textwidth]{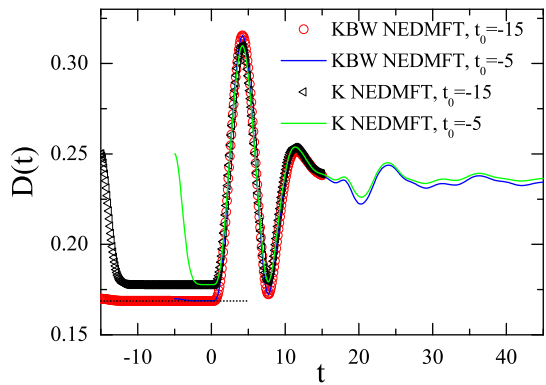}
\caption{(Color online) The time dependence of the double occupation 
$D(t)$ calculated within the Kadanoff-Baym-Wagner (KBW) and the Keldysh (K) 
NEDMFT in the weak interaction case for different initial times $t_0$.
The results are already scaled by a quadratic extrapolation with 
$\Delta t=0.1$, $0.065$, and $0.5$. The dotted line is the double occupation
in equilibrium ($E=0$).
The model parameters
$U=0.5$, $E=1$, $\Delta \tau=0.1$, $\beta=10$.}
\label{fig15}
\end{figure}

\begin{figure}[t]
\vspace{-1cm}
\includegraphics[width=0.5\textwidth]{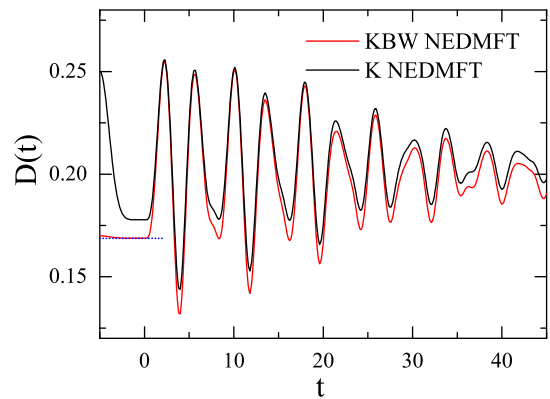}
\caption{(Color online) The time dependence of the double occupation 
$D(t)$ calculated within the Kadanoff-Baym-Wagner (KBW) and the Keldysh (K) 
NEDMFT in the weak interaction case for different initial times $t_0$.
The results are already scaled by a quadratic extrapolation with 
$\Delta t=0.1$, $0.065$, and $0.5$. The dotted line is the double occupation
in equilibrium ($E=0$).
The model parameters
$U=0.5$, $E=2$, $\Delta \tau=0.1$, $\beta=10$.}
\label{fig16}
\end{figure}

We also calculate the double occupation 
$D(t)=\langle c^{\dagger}(t)c(t) f^{\dagger}(t) f(t) \rangle$ which can be computed
through the lesser Green function $Q_1^{<}(t,t)$ defined in Eq.~(\ref{eqq})
by
\begin{eqnarray}
D(t) = - i n_f Q_1^{<}(t,t) .
\end{eqnarray}
In Fig.~\ref{fig15} we plot the time dependence of the double occupation $D(t)$
calculated within the Kadanoff-Baym-Wagner and the Keldysh NEDMFT in the weak
interaction case. Before the electric field is turned on ($t<0$), the double
occupation calculated within the Kadanoff-Baym-Wagner NEDMFT is constant in
a good agreement with the equilibrium value obtained by solving the DMFT equations
in frequency. However, the Keldysh NEDMFT results are quite different. Within
the Keldysh NEDMFT, the double occupation starts from its noninteraction value
at half filling ($D_0=0.25$), and then decreases to a steady value. This steady
value is little larger than the equilibrium value. At the initial time 
$t_0$ the Keldysh formalism starts with noninteracting system and electron 
correlations are absent. The results show that
before the electric field is turned on, the Keldysh formalism cannot restore 
full electron correlations of the system. 
The full electron correlations are essentially elaborated from the initial correlations
which come from the dynamics of the system in the imaginary time branch of the
Kadanoff-Baym contour. The neglect of the initial correlations also means the lack
of the electron correlations even when the system is still in equilibrium. In
Fig.~\ref{fig15} we also plot the double occupation for different initial times $t_0$.
It shows that even when the initial time goes to the  remote past, the lack of electron correlations
still occurs in the Keldysh formalism. Only in the Kadanoff-Baym formalism when
the initial correlations are taken into account, full electron correlations
are obtained. After the electric field is turned on ($t>0$) for weak and strong
electric fields (for instance, $E=1$) the double occupation
first oscillates strongly, and then is damped into  less regular oscillations.  
However, when the electric field increases further (for instance, $E=2$) the double occupation 
regularly oscillates even at
long time, as shown in Fig.~\ref{fig16}. This behavior is reminiscent to the beats of the electric
current. In the weak interaction case
the difference of the double occupations calculated within the Kadanoff-Baym-Wagner
and the Keldysh formalisms is small in relation with their values. After the turning on 
of the electric field  the difference becomes smaller. 
In the weak interaction case the Keldysh formalism qualitatively 
describes the behavior of the double occupation. 

\begin{figure}[t]
\vspace{-1cm}
\includegraphics[width=0.5\textwidth]{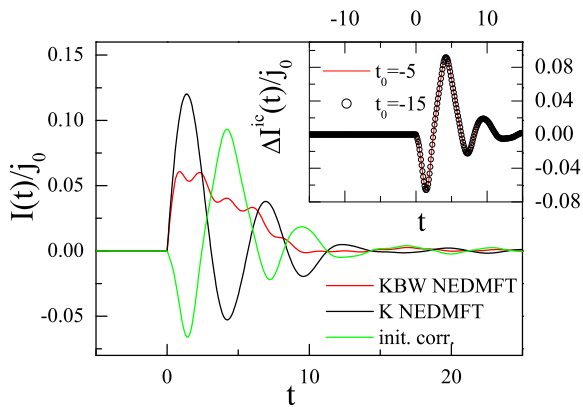}
\caption{(Color online) The time dependence of the electric current $I(t)/j_0$ calculated
within the Kadanoff-Baym-Wagner (KBW) and the Keldysh (K) NEDMFT 
for strong interaction $U=2$, $E=1$. The contribution of the initial correlations
is also plotted. The inset plots the initial correlation contribution for
different initial times $t_0$. The results are already scaled by a cubic
extrapolation with $\Delta t=0.05$, $0.035$, $0.025$, and $0.02$ 
( $\Delta \tau=0.1$, $\beta=10$).}
\label{fig17}
\end{figure}

In equilibrium when the interaction $U>\sqrt{2}$ the density of states opens
a gap at the Fermi energy, and the system is insulator.
It distinguishes between the weak and strong interaction cases. In general, in the
strong interaction case the numerical calculations slowly converge with $\Delta t$.
Usually, we have to use more small values of $\Delta t$ in order to get reliable results.
In Fig.~\ref{fig17} we plot the electric current in the strong interaction case.  
In contrast to the weak interaction case, the current does not display the regular Bloch
oscillations. The current oscillations are rather irregular and quenched. However, the
current calculated within the Keldysh NEDMFT still exhibits the regular Bloch oscillations. 
This shows that the initial correlations are important in the strong interaction case. They are
a main factor for quenching the current oscillations. The contribution of the initial correlations
to the current is not small as in the weak interaction case. It is of order of the current.
In the inset of Fig.~\ref{fig17} we also plot the initial correlation contribution to the current
for different initial times $t_0$. It shows that the contribution remains the same
as the initial time goes to the remote past.
Thus, in the strong interaction case the initial correlations become significant, and dominate
the overall properties of the current. The neglect of the initial correlations may cause artifacts in
the nonequilibrium properties of the current.

\begin{figure}[t]
\vspace{-1cm}
\includegraphics[width=0.5\textwidth]{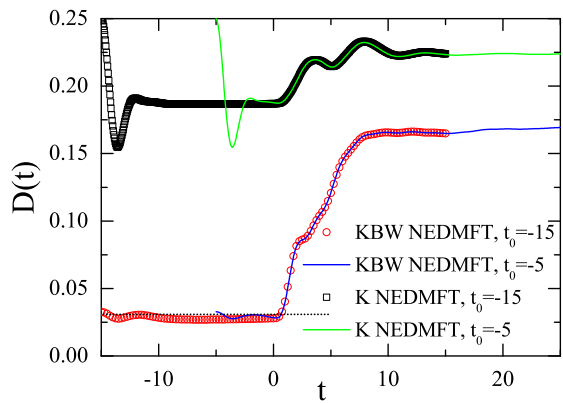}
\caption{(Color online) The time dependence of the double occupation $D(t)$ calculated
within the Kadanoff-Baym-Wagner (KBW) and the Keldysh (K) NEDMFT in the strong interaction case
for different initial times $t_0$.  The results are already scaled by a cubic
extrapolation with $\Delta t=0.05$, $0.035$, $0.025$, and $0.02$. The dotted line is
the double occupation in equilibrium ($E=0$). The model parameters 
$U=2$, $E=1$, $\Delta \tau=0.1$, $\beta=10$.}
\label{fig18}
\end{figure}

In Fig.~\ref{fig18} we plot the double occupation in the strong interaction case.
Before the turning on of the electric field, the time dependence of the double occupation
within both the Kadanoff-Baym-Wagner and the Keldysh NEDMFT is similar to the
weak interaction case. Within the Kadanoff-Baym-Wagner formalism the double occupation
is constant for $t<0$. The constant value is in a good agreement with the equilibrium
value, although there are very little deviations due to the finite size effects in
the numerical calculations. The double occupation calculated within the Keldysh
formalism first starts with the noninteraction value $D_0=0.25$ at the initial 
time $t_0$, then relaxes to a steady value. Like in the weak interaction case,
the steady value is not the equilibrium value. It again indicates that the Keldysh formalism
losses a some part of electron correlations. In the strong interaction case this lack of electron
correlations 
becomes significant. As a consequence, after the turning on of the electric field
the lack of electron correlations also remains significant. Due to the quenching of the
Bloch oscillations in the strong interaction case, the double occupation reaches
a steady value at a long time. The steady values obtained within the
Kadanoff-Baym-Wagner and the Keldysh formalisms are quite different. They indicate
the important contribution of the initial correlations. As shown in  Fig.~\ref{fig18}
the results do not change when the initial time goes to the remote past.   
For strong interactions the Keldysh formalism losses a significant part of electron correlations
both before and after the turning on of the electric field. It cannot correctly describe the nonequilibrium
properties.

\section{Conclusions}
In this paper we present the Kadanoff-Baym-Wagner formalism for nonequilibrium systems.
The formalism is based on the Wagner representation of the contour-ordered
Green function. Within the Kadanoff-Baym-Wagner formalism the Green functions
satisfy the proper boundary conditions. The initial correlations essentially distinguish between
the Kadanoff-Baym-Wagner and the Keldysh formalisms. We derive the system of equations
for nonequilibrium Green functions, and solve it within the truncated and
self-consistent perturbation theories as well as within the NEDMFT.
As a benchmark we examine the equilibrium FKM by the Kadanoff-Baym-Wagner NEDMFT.
The results show a good agreement between the Kadanoff-Baym-Wagner NEDMFT in equilibrium
and the equilibrium DMFT. In the nonequilibrium case the Green functions obtained 
within the Kadanoff-Baym-Wagner NEDMFT satisfy the spectral sum rule well.
The derived Kadanoff-Baym-Wagner equations for nonequilibrium Green functions
are an alternative useful method for studying nonequilibrium systems.

In this paper we also emphasize the initial correlations. Within the perturbation
theory the initial correlations always finite even when the initial time goes to the remote past.
The electric current calculated within the truncated perturbation theory shows that 
the Kadanoff-Baym-Wagner formalism overestimates the current, whereas the Keldysh
formalism underestimates it. However, the Kadanoff-Baym-Wagner perturbation
theory shows a better agreement with the exact solution. 
For a long time the truncated perturbation theory fails to describe the physical properties. 
The self-consistent perturbation theory gives better results than the
truncated perturbation theory. The time domain in which 
the self-consistent
perturbation theory gives reasonable results  is wider than the one in the truncated perturbation theory. 
However,
the self-consistent perturbation theory cannot reproduce the beat behaviors of the
current for very strong electric fields. 
Within the perturbation theories the initial correlations do not qualitatively change
the perturbation results. Since both the Kadanoff-Baym-Wagner and the Keldysh perturbation theory 
results are close, use of the Keldysh approach is more convenient since its equations are simpler.
In the infinite dimension limit the NEDMFT
gives the exact solution. Examining the NEDMFT within both the Kadanoff-Baym-Wagner
and the Keldysh formalisms one can figure out the role of the initial correlations.
For weak interactions the initial correlations give only small contributions to
the physical quantities such as the electric current or the double occupation.
However, they remain finite for the long time limit. The initial correlations
are also important even before the electric field is turned on when the system
is still in equilibrium. Without the initial correlations the system cannot restore
the full electron correlations. For strong interactions the initial correlations
become significant, and dominate the physical properties. Without taking into account the initial
correlations the Keldysh formalism
can qualitatively describe the nonequilibrium properties of the system only for weak interactions.
For strong interactions it fails to count full electron correlations.
The neglect of the initial correlations may cause artifacts in the nonequilibrium properties of the system. 

\begin{acknowledgments}

The author would like to thank the Asia Pacific Center
for Theoretical Physics for the hospitality where the main
part of this work was done.
He also acknowledges
useful discussions with Han-Yon Choi, and thanks J.~K.~Freericks and V.~Turkowski
for providing their numerical data.
The author is grateful to thank the Max Planck Institute
for the Physics of Complex Systems at Dresden for sharing computer facilities
where the numerical calculations were performed.
This work was supported by
the Asia Pacific Center for Theoretical Physics, and in part by the Vietnam 
National Program on Basic Research.

\end{acknowledgments}

\end{document}